\documentclass[sn-mathphys,Numbered]{sn-jnl}

\usepackage{graphicx}%
\usepackage{multirow}%
\usepackage{amsmath,amssymb,amsfonts}%
\usepackage{amsthm}%
\usepackage{mathrsfs}%
\usepackage[title]{appendix}%
\usepackage{xcolor}%
\usepackage{textcomp}%
\usepackage{manyfoot}%
\usepackage{booktabs}%
\usepackage{algorithm}%
\usepackage{algorithmicx}%
\usepackage{algpseudocode}%
\usepackage{listings}%
\usepackage{multirow}%
\usepackage{tabularx}%

\usepackage{bm}
\DeclareMathOperator*{\argmax}{arg\,max}
\DeclareMathOperator*{\argmin}{arg\,min}
\newcommand{\gevmax}[2]{{\rm GEV_{max}}\left(#1,#2\right)}
\newcommand{\gevmin}[2]{{\rm GEV_{min}}\left(#1,#2\right)}
\newcommand{\sevmax}[1]{{\rm SEV_{max}}\left(#1\right)}

\newcommand{\vecw}{\bm{w}}
\newcommand{\vecs}{\bm{s}}
\newcommand{\vecn}{\bm{n}}
\newcommand{\vecx}{\bm{x}}
\newcommand{\vece}{\bm{e}}
\newcommand{\vech}{\bm{h}}
\newcommand{\phix}{\bm{\Phi}_{\rm x}}
\newcommand{\phis}{\hat{\bm{\Phi}}_{\rm s}}
\newcommand{\phin}{\hat{\bm{\Phi}}_{\rm n}}
\newcommand{\phisigma}{\bm{\Phi}_{\rm \sigma}}
\newcommand{\phir}{\bm{\Phi}_{\rm r}}
\newcommand{\ms}{m_{\rm s}}
\newcommand{\mn}{m_{\rm n}}
\newcommand{\maskp}{m_{\rm p}}
\newcommand{\maskset}{\mathcal M}

\newcommand{\htp}[1]{{#1}^{\rm H}}
\newcommand{\tp}[1]{{#1}^{\rm T}}
\newcommand{\conj}[1]{\overline{#1}}
\newcommand{\average}[1]{\left\langle#1\right\rangle}

\newcommand{\reffig}[1]{Fig.~\ref{#1}}
\newcommand{\reftable}[1]{Table~\ref{#1}}
\newcommand{\refeqn}[1]{(\ref{#1})}
\newcommand{\refsec}[1]{Section~\ref{#1}}
\newcommand{\refsubsec}[1]{{\bf\ref{#1}}}
\newcommand{\nop}[1]{}

\newcolumntype{C}[1]{>{\centering\arraybackslash}p{#1}}
\newcolumntype{L}[1]{>{\raggedright\arraybackslash}p{#1}}
\newcolumntype{R}[1]{>{\raggedleft\arraybackslash}p{#1}}

\begin{document}

\title[Can all variations within the unified mask-based beamformer framework achieve identical peak extraction performance?]
{Can all variations within the unified mask-based beamformer framework achieve identical peak extraction performance?}


\author*[1]{\fnm{Atsuo} \sur{Hiroe}}\email{hiroe@ra.sc.e.titech.ac.jp}
\author[2]{\fnm{Katsutoshi} \sur{Itoyama}}\email{katsutoshi.itoyama@jp.honda-ri.com}
\author[1]{\fnm{Kazuhiro} \sur{Nakadai}}\email{nakadai@ra.sc.e.titech.ac.jp}
\affil*[1]{Department of Systems and Control Engineering, School of Engineering, Institute of Science Tokyo, \city{Tokyo}, \country{Japan}}
\affil*[2]{Honda Research Institute Japan Co., Ltd., \city{Saitama}, \country{Japan}}



\abstract{%
This study investigates mask-based beamformers (BFs), which estimate filters for target sound extraction (TSE) using time-frequency masks. Although multiple mask-based BFs have been proposed, no consensus has been reached on which one offers the best target-extraction performance. Previously, we found that maximum signal-to-noise ratio and minimum mean square error (MSE) BFs can achieve the same extraction performance as the theoretical upper-bound performance, with each BF containing a different optimal mask. However, two issues remained unsolved: only two BFs were covered, excluding the minimum variance distortionless response BF; and ideal scaling (IS) was employed to ideally adjust the output scale, which is not applicable to realistic scenarios. To address these issues, this study proposes a unified framework for mask-based BFs comprising two processes: filter estimation that can cover all possible BFs and scaling applicable to realistic scenarios by employing a mask to generate a scaling reference. Based on the operators and covariance matrices used in BF formulas, all possible BFs can be classified into 12 variations, including two new ones. Optimal masks for both processes are obtained by minimizing the MSE between the target and BF output. The experimental results using the CHiME-4 dataset suggested that 1)~all 12 variations can achieve the theoretical upper-bound performance, and 2)~mask-based scaling can behave like IS, even when constraining the temporal mean of a non-negative mask to one. These results can be explained by considering the practical parameter count of the masks. These findings contribute to 1)~designing a TSE system, 2)~improving scaling accuracy through mask-based scaling, and 3)~estimating the extraction performance of a BF.}




\maketitle

\section{Introduction}\label{sec:intro}

Target sound extraction (TSE) estimates a sound source of interest, namely the target, from mixtures of multiple sources. This is effective in improving speech intelligibility in telecommunication systems and the performance of automatic speech recognition (ASR) systems~\cite{Chen2018-rl,Zmolikova2023-ku}. Beamformers (BFs) are employed as a linear TSE method to avoid nonlinear distortions such as musical noises and spectral distortions~\cite{,Zmolikova2023-ku, Swietojanski2014-sa, Mizumachi2016-jd, Mizumachi2019-hi}. In the last decade, combined frameworks comprising BFs and deep neural networks (DNNs), referred to as mask-based BFs, have been proposed~\cite{Heymann2016-eb, Heymann2016-sy, Erdogan2016-jc}. In these frameworks, DNNs generate one or two time-frequency (TF) masks corresponding to the target, interferences, or both to inform the BF of the sound to be enhanced or suppressed. Subsequently, the BF estimates a filter for extracting the target using these masks. For filter estimation, the following BF types are adopted: 1)~maximum signal-to-noise ratio (max-SNR) or generalized eigenvalue (GEV) BF~\cite{Heymann2016-eb, Heymann2016-sy, Drude2019-so}, 2)~minimum variance distortionless response (MVDR) BF~\cite{Heymann2016-sy, Erdogan2016-jc, Souden2010-kp}, and 3)~minimum mean square error (MMSE) or multichannel Wiener filter (MWF) BF~\cite{Stenzel2013-jm, Nugraha2016-rn, Pfeifenberger2017-zp}.

Our interest is to determine which BF type can achieve the best extraction performance in estimating the target sound. Although several studies have compared multiple types~\cite{Heymann2016-sy,Boeddeker2018-ww,Wang2018-tl,Shimada2019-xj,Heymann2018-mf}, no consensus has been established; some found the max-SNR BF to be the best~\cite{Heymann2016-sy}, whereas others favored the MMSE BF~\cite{Wang2018-tl,Shimada2019-xj} and MVDR BF~\cite{Boeddeker2018-ww}. In another study, performance depended on the number of microphones used~\cite{Heymann2018-mf}. Moreover, no consensus has been established for the best mask type although different mask types such as binary, ratio, and complex-valued masks have been examined to train the mask-estimating DNNs~\cite{Heymann2016-eb,Erdogan2016-jc,Xu2019-um}. Therefore, we are motivated to explore the best BF and mask type under the same conditions and independent of DNNs. This is a significant preliminary stage for designing the best TSE system using a mask-based BF.

As the first step of this stage, our previous study~\cite{Hiroe2023-qh} compared four BFs: the max-SNR BF, its two variations that use a single mask, and MMSE BF under unified conditions. We used the CHiME-3 simulated test set~\cite{Barker2017-tr} and obtained the optimal mask for each utterance by minimizing the mean square error (MSE) between the BF output and target clean speech. Ideal scaling (IS) was employed as the unified scaling (or post-filtering) method to adjust the scale of the BF output in each frequency bin. The source-to-distortion ratio (SDR) was measured as the evaluation score. Consequently, we obtained the following findings: 
\begin{enumerate}
    \item All four BFs can achieve the same peak performance, comparable with the theoretical upper-bound performance obtained with the ideal MMSE.
    \item The optimal mask is unique for each BF method.
    \item The ideal mask for the single-channel masking differs from the optimal mask for the mask-based BFs.
\end{enumerate}
Considering that the aforementioned comparative studies~\cite{Heymann2016-sy,Boeddeker2018-ww,Wang2018-tl,Shimada2019-xj,Heymann2018-mf} were based on the intuition that the optimal mask should be common for any BF and that the peak performance achieved with the mask should differ for each BF, our findings are contrary to this. However, achieving the best TSE system leveraging these findings presents two challenges: 1)~all BF types should be covered, and 2)~a scaling method free from the target sound is required.

First, our previous study only examined four BFs derived from two types, namely the max-SNR and MMSE BFs. The MVDR BF, although extensively employed, was not examined. Moreover, multiple variations can be derived within each BF type. Therefore, a framework that covers all possible variations should be established rather than simply examining existing BFs one by one. 

Second, our previous study employed the IS because a common scaling method independent of the BF type was required to verify whether the BFs can achieve the same performance. However, the IS is not applicable to realistic scenarios because it requires the target sound as a scaling reference. Therefore, we need an alternative scaling method independent of the BF used, free from the target sound, and comparable to IS in scaling performance. 

Reflecting on these aspects, we propose a unified framework for mask-based BFs. This framework consists of two mask-based processes: filter estimation and scaling. The former process can cover all variations, and the latter is free from the target and independent of the BF variation used. Additionally, we employ a classification rule based on the operators and covariance matrices included in each BF formula to enumerate all possible variations of the mask-based BFs. According to this rule, 12 variations, including two novel ones, can be identified in total. Using this framework, we can rephrase our interest as follows: 1)~whether all possible variations can achieve the upper-bound extraction performance, 2)~whether the mask-based scaling is comparable to the IS, and 3) which mask type (or constraint) is the best for each process. This study experimentally verifies these aspects by obtaining the optimal masks and discussing the reasons for the experimental results. 

Through enumerating all possible variations, we found that the formulas of several variations are also employed in another type of linear TSE based on the independent component analysis (ICA) theory~\cite{Hiroe2022-rd,Cho2019-pn,Shin_undated-za}, referred to as ICA-based TSE, although these are derived from a different formulation from mask-based BFs. Therefore, this study treats these TSE methods as BF variations and considers that the insights obtained from the experimental results apply to the methods.

This study contributes to the following aspects: 1)~the unified framework facilitates designing the best TSE system using BFs; 2)~the mask-based scaling combined with any BF can improve the scaling accuracy; 3)~the discussion based on the practical parameter count and saturation point can estimate the peak extraction performance of the BF used.

The remainder of this paper is organized as follows.
Section \ref{sec:overview-mask-based-bfs} overviews existing mask-based BFs. \refsec{sec:unified-framework} proposes a unified framework for mask-based BFs. \refsec{sec:experiments} experimentally verifies the aforementioned aspects, while \refsec{sec:discussion} discusses the experimental results. Finally, \refsec{sec:conclusions} concludes the study.

\section{Overview of mask-based BFs} \label{sec:overview-mask-based-bfs}

Given that this study examines all possible mask-based BFs, this section provides an overview of existing ones. First, we discuss peak extraction performance and the concept of the optimal mask. After introducing the signals used, we enumerate all existing variations of mask-based BFs,including ICA-based TSE methods. Finally, we examine the mask types and scaling methods in each subsection.

\subsection{What is the peak extraction performance and optimal mask?} \label{subsec:def_performance}
\begin{figure}[t]
  \centering
  \includegraphics[width=0.4\linewidth]{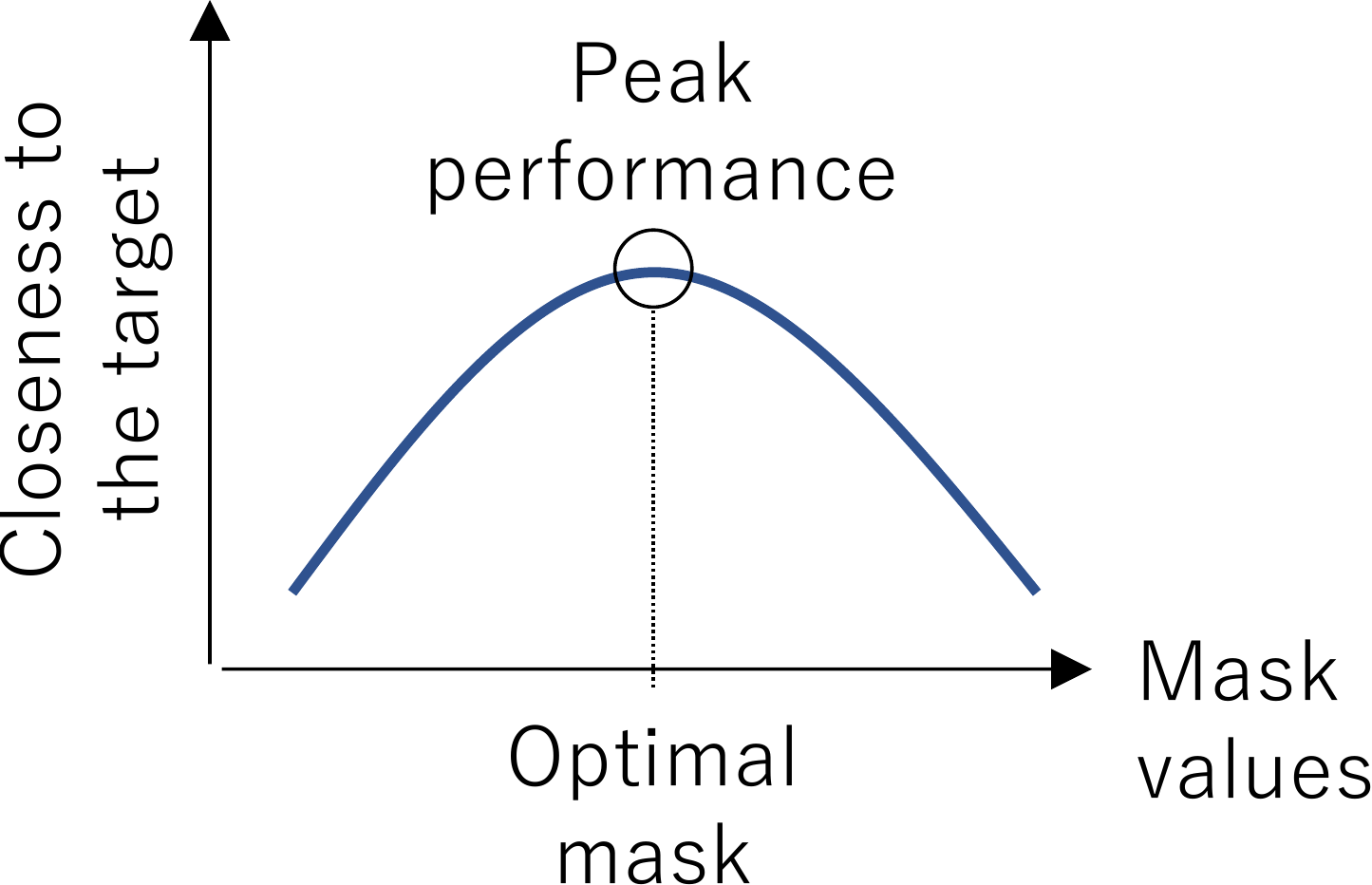}
  \caption{Conceptual plot of the relationship between the closeness of the BF output to the target and mask values; the optimal mask denotes a set of mask values that achieve the BF output closest to the target.}
  \label{fig:mask_curve}
\end{figure}

In this study, extraction performance is considered to be the BF output closest to the target in the TF domain, given that a significant goal of BFs is to extract (or estimate) the target. The peak performance and optimal mask are explained in \reffig{fig:mask_curve}; the vertical and horizontal axes indicate the closeness of the BF output to the target and mask values, respectively. Although mask values vary multidimensionally, this figure conceptually represents the variation as a single axis. The extraction performance depends on this variation and exhibits a peak at a particular mask value. We refer to this as the optimal mask. As mentioned in \refsec{sec:intro}, the optimal mask differs for each BF even when inputting the same observations.

\subsection{Signal models}\label{subsec:signal-models}
This study considers that all signals are in the TF domain. The frequency index is omitted for simplicity, whereas the frame index $t$ is always described. Let $\vecx(t)=\tp{[x_{1}(t),\ldots ,x_{N}(t)]}$ be an observation vector obtained with $N$ microphones. The observation $\vecx(t)$ can be expressed as the following mixture:
\begin{equation}
    \vecx(t) = \vecs(t) + \vecn(t), \label{eqn:components-of-x}
\end{equation}
where $\vecs(t)=\tp{[s_1(1), \ldots, s_N(t)]}$ denotes the components arriving from the target source and $\vecn(t)=\tp{[n_1(1), \ldots, n_N(t)]}$ represents the residuals called interferences. Using the observation $\vecx(t)$ and extraction filter $\vecw$, the estimated target $y(t)$ is expressed as
\begin{equation}
    y(t) = \htp{\vecw} \vecx(t). \label{eqn:bf-output}
\end{equation}
Several BF types require scaling $y(t)$ as a post-process. The scaling process can be represented as
\begin{eqnarray}
z(t) &=& \gamma y(t), \label{eqn:scaling-def}
\end{eqnarray}
where $z(t)$ and $\gamma$ are referred to as the {\it BF output} and {\it scaling factor}, respectively. This study considers that the process adjusts not only the magnitude of $y(t)$ but also its phase. Thus, $\gamma$ should be not real-valued but complex-valued for more accurate scaling.

To estimate $\vecw$, we define the following covariance matrices:
\begin{eqnarray}
\phix &=& \average{\vecx(t)\htp{\vecx(t)}}_{t},
    \label{eqn:phix}\\
\phis &=& \average{\ms(t)\vecx(t)\htp{\vecx(t)}}_{t},
    \label{eqn:phis}\\
\phin &=& \average{\mn(t)\vecx(t)\htp{\vecx(t)}}_{t},
    \label{eqn:phin}
\end{eqnarray}
where $\ms (t)$ and $\mn (t)$ denote TF masks for the target and interferences, respectively, and $\average{\cdot}_{t}$ computes the average over $t$. We refer to $\phix$, $\phis$, and $\phin$ as observation, target, and interference covariance matrices, respectively. Unlike $\phix$, both $\phis$ and $\phin$ are estimated matrices computed from the masks and observations without using $\vecs(t)$ and $\vecn(t)$. Constraints to the mask values are mentioned in \refsubsec{subsec:mask-types}.

We consider that the optimal mask is the solution to the following minimization problem:
\begin{equation}
\maskset_{\rm filt} = \argmin_{\maskset_{\rm filt}} \average{\left|s_{k}(t) - z(t)\right|^{2}}_{t},
\label{eqn:optimal-mask-def}
\end{equation}
where $k$ is the reference microphone index, and $\maskset_{\rm filt}$ denotes a set of mask values that comprises $\ms(t)$, $\mn(t)$, or both for all $t$, depending on the BF employed. In principle, $\maskset_{\rm filt}$ cannot be obtained as the closed-form solution because the masks are indirectly used to estimate $\vecw$ in \refeqn{eqn:bf-output}. 

We refer to the eigenvectors corresponding to the maximum and minimum eigenvalues simply as the {\it maximum} and {\it minimum} eigenvectors, respectively. Then, consider $\gevmax{\bm{A}}{\bm{B}}$ and $\gevmin{\bm{A}}{\bm{B}}$ to be the maximum and minimum eigenvectors in the GEV problem represented as \refeqn{eqn:gev}, respectively. Similarly, consider $\sevmax{\bm{A}}$ to be the maximum eigenvector in the standard eigenvector (SEV) problem represented as \refeqn{eqn:sev}.
\begin{eqnarray}
    \bm{Aw} &=& \lambda \bm{Bw} \label{eqn:gev}\\
    \bm{Aw} &=& \lambda \bm{w} \label{eqn:sev}
\end{eqnarray}

We also use $\vech=[h_{1},\ldots,h_{N}]$, $\vece_{k}$, ${\rm tr}(\cdot)$, and $\max(\cdot)$ as the steering vector (SV) corresponding to the target sound direction, one-hot vector in which the only $k$th element is one whereas the others are zero, trace of the given matrix, and the maximum value among the given arguments, respectively.

\subsection{Formulas used in existing mask-based BFs}\label{subsec:formulas-mask-based-bfs}

This study compares all possible variations of the mask-based BFs regardless of whether they have been employed. Thus, we enumerate them by examining the formulations of the max-SNR, MMSE, and MVDR BFs in Appendices \ref{subsec:max-snr}, \ref{subsec:mmse}, and \ref{subsec:mvdr}, respectively. \reftable{tab:formulas-mask-based-bfs} shows the derived formulas, indicating whether each variation contains the scaling ambiguity issue that scales of $\vecw$ and $y(t)$ are undetermined.

\begin{table}[b]
  \centering
\caption{Formulas used in existing mask-based BFs}
\label{tab:formulas-mask-based-bfs}
\begin{tabular*}{\textwidth}{@{\extracolsep\fill}p{0.55\textwidth}lC{5em}}
\toprule
BF name & Formula for filter $\vecw$& Scaling ambiguity\\
\midrule\midrule
Maximum signal-to-noise ratio (max-SNR) \cite{Heymann2016-eb, Heymann2016-sy, Drude2019-so} 
&$\vecw=\gevmax{\phis}{\phin}$ &$\checkmark$\\  
Maximum observation-to-noise ratio (max-ONR) \cite{Warsitz2007-fn}
&$\vecw=\gevmax{\phix}{\phin}$ &$\checkmark$\\  
Maximum signal-to-observation ratio (max-SOR) \cite{Hiroe2023-qh}
&$\vecw=\gevmax{\phis}{\phix}$ &$\checkmark$\\  
Minimum noise-to-signal ratio (min-NSR) 
&$\vecw=\gevmin{\phin}{\phis}$ &$\checkmark$\\  
Minimum noise-to-observation ratio (min-NOR) \cite{Hiroe2022-rd,Hiroe2023-qh}
&$\vecw=\gevmin{\phin}{\phix}$ &$\checkmark$\\  
Minimum observation-to-signal ratio (min-OSR) 
&$\vecw=\gevmin{\phix}{\phis}$ &$\checkmark$\\  
\midrule
Minimum mean square error (MMSE) \cite{Stenzel2013-jm, Nugraha2016-rn, Pfeifenberger2017-zp}
&$\vecw=\phix^{-1} \phis \vece_{k}$ & \\
\midrule
Minimum variance distortionless response (MVDR) \cite{Heymann2016-sy, Erdogan2016-jc, Boeddeker2017-tu}
&$\vecw=\dfrac{\phin^{-1}\vech}{\htp{\vech}\phin^{-1}\vech}$ & $\checkmark$\footnotemark[1]\\
Minimum power distortionless response (MPDR) \cite{Ehrenberg2010-tq}
&$\vecw=\dfrac{\phix^{-1}\vech}{\htp{\vech}\phix^{-1}\vech}$ & $\checkmark$\footnotemark[1]\\
Souden MVDR \cite{Souden2010-kp}
&$\vecw=\dfrac{\phin^{-1}\phis\vece_{k}}{{\rm tr}\left(\phin^{-1}\phis\right)}$ &\\
\bottomrule
\end{tabular*}
\footnotetext[1]{In the case that the norm of $\vech$ is undetermined}
\end{table}

The first to sixth rows denote variations of the max-SNR BF. These commonly contain the scaling ambiguity issue. The min-NSR, min-NOR, and min-OSR BFs are equivalent to max-SNR, max-ONR, and mas-SOR, respectively as mentioned in Appendix \ref{subsec:max-snr}.

The seventh row denotes the MMSE BF. This BF can determine the output scale. This also indicates that the range of mask values
$\ms(t)$ are sensitive to both the magnitude and phase of $y(t)$.

The eighth to tenth rows denote variations of the MVDR BF. For both the MDVR and MPDR, the SV $\vech$ can be computed as the maximum eigenvector of $\phis$~\cite{Heymann2016-sy,Boeddeker2017-tu}:
\begin{equation}
\vech = \sevmax{\phis}.
    \label{eqn:steering-vector}
\end{equation}
Considering that the eigenvalue problem cannot determine the eigenvector norms, the two BFs contain the scaling ambiguity issue. Contrary, the Souden MVDR BF is free from the issue because this does not employ $\vech$.

This study also employs the ideal MMSE BF~\cite{Malek2020-nw}, which can achieve the theoretical upper-bound extraction performance for all BFs by minimizing the MSE between $y(t)$ and the target; when $\vecs(t)$ in \refeqn{eqn:components-of-x} is known, the ideal filter can be obtained using an element of $\vecs(t)$ as the ideal reference:
\begin{eqnarray}
\vecw_{\rm ideal} &=& \argmin_{\vecw} \average{\left|s_{k}(t) - y(t)\right|^{2}}_{t}
    \label{eqn:ideal-mmse-def} \\
                  &=& \phix^{-1} \average{\vecx(t) \conj{s_{k}(t)}}
                  \quad \left(=\phix^{-1} \average{\vecx(t) \htp{\vecs(t)}}_{t}\vece_{k}\right),
    \label{eqn:ideal-mmse}
\end{eqnarray}
where $\conj{s_{k}(t)}$ denotes the conjugate of $s_{k}(t)$.

Our previous study examined the max-SNR, max-SOR, min-NOR, and MMSE BFs and found the following aspects:
\begin{enumerate}
\item The four BFs achieve the same extraction performance comparable with the ideal MMSE BF.
\item The optimal mask is unique for each BF. For example, the optimal masks for the max-SNR BF ($\ms(t)$ and $\mn(t)$) are not optimal for the max-SOR or min-NOR BFs. Similarly, the optimal mask for the MMSE BF is not optimal for the max-SOR BF, and vice versa.
\item The four BFs using the ideal ratio mask~\cite{Wang2018-yy} are not comparable with the ideal MMSE BF.
\end{enumerate}

\begin{table}[t]
  \centering
\caption{Formulas used in ICA-based TSE methods; SIBF and MLDR correspond to min-NOR and MVDR BFs respectively. ($r(t)$: reference that is estimated magnitude spectrogram of the target, $\beta$: reference exponent that controls the influence of $r(t)$, $\varepsilon$: threshold that prevents zero-division, $\sigma(t)^{2}$: time-frequency-varying variance)}
\label{tab:formulas-ica-based-tse}
\begin{tabular*}{\textwidth}
{@{\extracolsep\fill}p{0.35\textwidth}p{0.33\textwidth}p{0.25\textwidth}}
\toprule
Name & Formula for filter & Corresponding to\\
\midrule\midrule
Similarity-and-independence-aware BF (SIBF) \cite{Hiroe2020-sy,Hiroe2022-rd}
&$\vecw=\gevmin{\phir}{\phix}$,\par where $\phir=\average{\dfrac{\vecx(t)\htp{\vecx(t)}}{\max(r(t)^{\beta},\varepsilon)}}_{t}$ & Min-NOR \par $\left(\mn(t)=\dfrac{1}{\max(r(t)^{\beta},\varepsilon)}\right)$\\
\midrule
Maximum likelihood distortion-less response (MLDR) BF \cite{Cho2019-pn,Cho2021-ru,Shin_undated-za}
& $\vecw=\dfrac{\phisigma^{-1}\vech}{\htp{\vech}\phisigma^{-1}\vech}$, \par where $\phisigma=\average{\dfrac{\vecx(t)\htp{\vecx(t)}}{\sigma(t)^{2}}}_{t}$ & MVDR \par $\left(\mn(t)=\dfrac{1}{\sigma(t)^{2}}\right)$\\
\bottomrule
\end{tabular*}
\end{table}

As mentioned in \refsec{sec:intro}, several ICA-based TSE methods employ the same formulas as those used in the mask-based BFs. One is the similarity-and-independence-aware BF (SIBF)~\cite{Hiroe2020-sy,Hiroe2022-rd}, and the other is the maximum likelihood distortion-less response (MLDR) BF~\cite{Cho2019-pn,Cho2021-ru,Shin_undated-za}. These assume that the target follows a particular distribution referred to as {\it source model} and obtain the extraction filter by maximizing the likelihood of the target under different constraints, as explained in Appendices \refsubsec{subsec:sibf} and \refsubsec{subsec:mldr}. \reftable{tab:formulas-ica-based-tse} shows the formulas used in both methods. Considering that the weighted matrices $\phir$ and $\phisigma$ correspond to $\phin$ for the mask-based BFs, the formulas of the SIBF and MLDR are identical to those of the min-NOR and MVDR BFs, respectively~\cite{Hiroe2022-rd,Cho2019-pn}. Therefore, we consider that the two methods are variations of the mask-based BFs and that the peak performance analysis of the mask-based BFs applies to the two.

\subsection{Mask types used in conventional mask-based BFs} \label{subsec:mask-types}
\begin{figure}[b]
  \centering
  \includegraphics[width=0.7\linewidth]{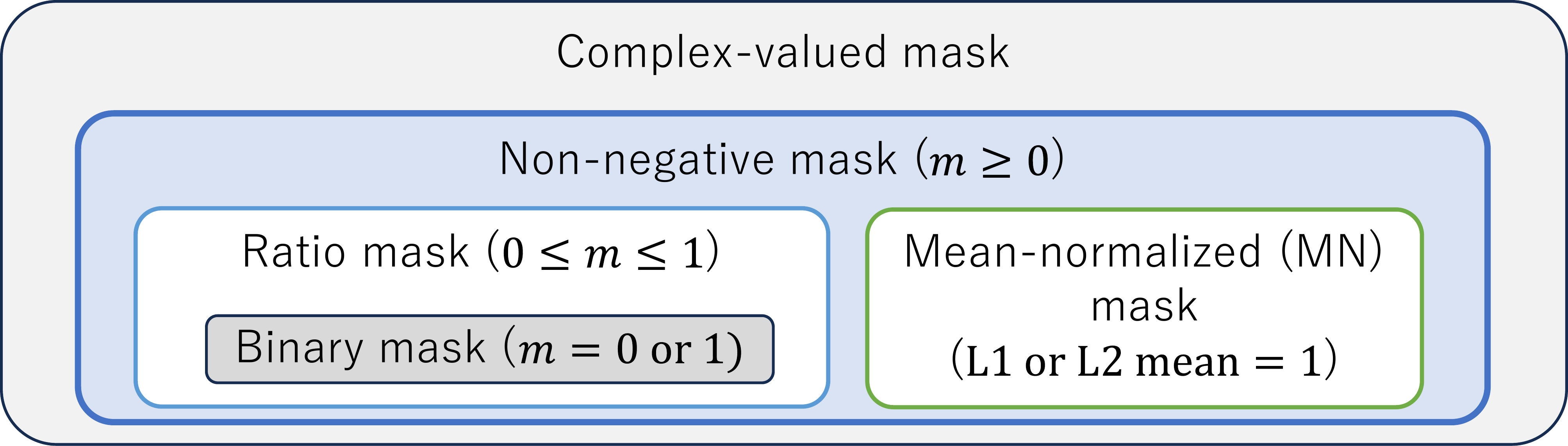}
  \caption{Mask type categorization based on the mask value constraint; the non-negative mask can be more constrained in two ways. One is a ratio mask and the other is a family of mean-normalized (MN) masks.}
  \label{fig:mask-categorization}
\end{figure}

We overview the mask types employed for the mask-based BFs. Considering the constraints on the mask values, we classify the types as illustrated in \reffig{fig:mask-categorization}. The complex-valued mask is the least constrained and can contain any complex numbers. Restricting the phase angle of the mask to 0 generates the non-negative mask. This mask can be more constrained in two ways. One is a ratio mask that limits its value to the range between 0 and 1. The binary mask is a particular case of this mask. The other is a family of mean-normalized (MN) masks that restrict their mean over $t$ to 1. Conventionally, two different constraints have been employed for this mask: L1-MN and L2-MN masks represented in \refeqn{eqn:l1-mean-normalized} and \refeqn{eqn:l2-mean-normalized}, respectively. 
\begin{eqnarray}
\average{m(t)}_{t}&=&1, \label{eqn:l1-mean-normalized}\\
\sqrt{\average{m(t)^{2}}_{t}}&=&1, \label{eqn:l2-mean-normalized}
\end{eqnarray}
where $m(t)$ denotes $\ms(t)$ or $\mn(t)$.

Investigating studies that employ DNNs to estimate masks for the BFs, we found that the following data need to be distinguished, although they can all be called masks:
\begin{itemize}
\item Supervisory data used for training the mask-estimating DNNs
\item DNN outputs
\item Weights used for computing weighted covariance matrices such as $\phis$ and $\phin$.
\end{itemize}
This study focuses on the third aspect. Note that computing $\phis$ and $\phin$ with \refeqn{eqn:phi-l1-mean-normalize} instead of \refeqn{eqn:phis} and \refeqn{eqn:phin} practically imposes the constraint represented as \refeqn{eqn:l1-mean-normalized} on the weights even when the DNN outputs are the ratio masks.
\begin{eqnarray}
\bm{\Phi} &=& \frac{\sum_{t}m(t)\vecx(t)\htp{\vecx(t)}}{\sum_{t}m(t)}, \label{eqn:phi-l1-mean-normalize}
\end{eqnarray}
where $\bm{\Phi}$ denotes $\phis$ or $\phin$. Thus, we consider that studies using \refeqn{eqn:phi-l1-mean-normalize} employ the L1-MN mask.

\begin{table}[t]
  \centering
\caption{Mask types used in mask-based BFs (MN: mean-normalized, JT: joint training, AM: acoustic model); this study focuses on weights for $\phis$ and $\phin$.}
\label{tab:mask-types}
\begin{tabular*}{\textwidth}
{@{\extracolsep\fill}lp{0.15\textwidth}p{0.21\textwidth}p{0.1\textwidth}p{0.15\textwidth}}
\toprule
       & BF & Supervisory data & DNN outputs & Weights for $\phis$ and $\phin$ \\
\midrule
Heymann+15~\cite{Heymann2016-eb}
& Max-SNR & Binary & Ratio & Ratio\\
Erdogan+16~\cite{Erdogan2016-jc}
& Souden MVDR & Other (target magnitude spectrogram) & Ratio & Ratio\\
Pfeifenberger+17~\cite{Pfeifenberger2017-zp}
& Max-SNR & Ratio & Ratio & L1-MN\\
Xu+19~\cite{Xu2019-um}
& Max-SNR and MVDR & n/a (JT with BF and AM) & Complex & L1-MN\\
Nguyen+22~\cite{Nguyen2022-hg}
& Souden MVDR & n/a (JT with BF) & Complex & Non-negative\\
\bottomrule
\end{tabular*}
\end{table}

\reftable{tab:mask-types} shows the mask types used in conventional studies. In~\cite{Xu2019-um} and \cite{Nguyen2022-hg}, no explicit supervisory data were provided for the mask-estimating DNNs because the DNNs were jointly trained with the downstream tasks including the BF. Significantly, the non-negative and more constrained masks were used as the weights of $\phis$ and $\phin$, regardless of the DNN outputs; in~\cite{Xu2019-um}, the complex-valued DNN outputs were converted to speech presence probabilities that can be interpreted as the L1-MN masks because both $\phis$ and $\phin$ were computed with \refeqn{eqn:phi-l1-mean-normalize}; in~\cite{Nguyen2022-hg}, the phases of the DNN outputs were ignored in computing $\phis$ and $\phin$. However, whether using these constrained masks degrades the BF extraction performance has not been investigated.

\subsection{Scaling methods} \label{subsec:scaling-methods}
Mask-based BFs other than the MMSE and Sounden MVDR BFs suffer from the scaling ambiguity issue mentioned in \refsubsec{subsec:formulas-mask-based-bfs}. Here, we overview scaling methods that adjust the scale of the BF output.

\reftable{tab:scaling-methods} shows conventional scaling methods combined with the mask-based BFs. The BAN and SWF calculate the scaling factor $\gamma$ within \refeqn{eqn:scaling-def}. These can be combined with the mask-based BFs that employ $\phin$. Note that both methods can only adjust the magnitude of the BF output because $\gamma$ is non-negative. In contrast, the MDP can adjust both the magnitude and the phase because $\gamma$ is complex-valued, and can be combined with any linear TSE methods including the SIBF. Unlike these methods, RTF modifies the SV $\vech$; thus, this can only be employed for the MVDR, MPDR, and MLDR BFs.

\begin{table}[t]
  \centering
\caption{Conventional scaling methods ($\hat{\sigma}_{s}^{2}$: estimated variance of the target)}
\label{tab:scaling-methods}
\begin{tabular*}{\textwidth}
{@{\extracolsep\fill}llC{5em}C{5em}}
\toprule
Name & Formula & Adjusting magnitude & Adjusting phase\\
\midrule\midrule
Blind analytical normalization (BAN) \cite{Warsitz2007-fn}&
$\gamma=\dfrac{\sqrt{\htp{\vecw}\phin\phin\vecw/N}}{\htp{\vecw}\phin\vecw}$ &$\checkmark$&\\
\midrule
Single-channel Wiener filter (SWF) \cite{Gannot2008-tv,Gannot2017-dt} &
$\gamma=\dfrac{\hat{\sigma}_{s}^{2}}{\hat{\sigma}_{s}^{2} + \htp{\vecw}\phin\vecw}$ &$\checkmark$&\\
\midrule
Minimal distortion principle (MDP) \cite{Matsuoka2002-yr}&
$\gamma=\dfrac{\average{x_{k} \conj{y(t)}}_{t}}{\average{\left|y(t)\right|^{2}}_{t}}$ &$\checkmark$&$\checkmark$\\
\midrule
Relative transfer function (RTF) \cite{Gannot2001-ni,Cohen2004-mn} &
Using $\vech/h_{k}$ instead of $\vech$ &$\checkmark$&$\checkmark$\\
\bottomrule
\end{tabular*}
\end{table}

Employing different scaling methods can cause inconsistency of the best BF mentioned in \refsec{sec:intro}. Thus, we previously used the IS \cite{Hiroe2023-qh} as a unified scaling method. This can obtain the best scaling factor in terms of MSE between the target $s_{k}(t)$ and the BF output $z(t)$ because this is formulated as follows:
\begin{eqnarray}
\gamma_{\rm ideal} &=& \argmin_{\gamma} \average{\left|s_{k}(t) - z(t)\right|^{2}}_{t} \label{eqn:ideal-scaling-def}\\
                   &=& \frac{\average{s_{k}(t)\conj{y(t)}}_{t}} {\average{\left|y(t)\right|^{2}}_{t}}.
        \label{eqn:ideal-scaling}
\end{eqnarray}
However, IS does not apply to realistic scenarios because $s_{k}(t)$ is unavailable.

\section{Unified framework for mask-based BFs} \label{sec:unified-framework}
\begin{figure}[t]
  \centering
  \includegraphics[width=1.0\linewidth]{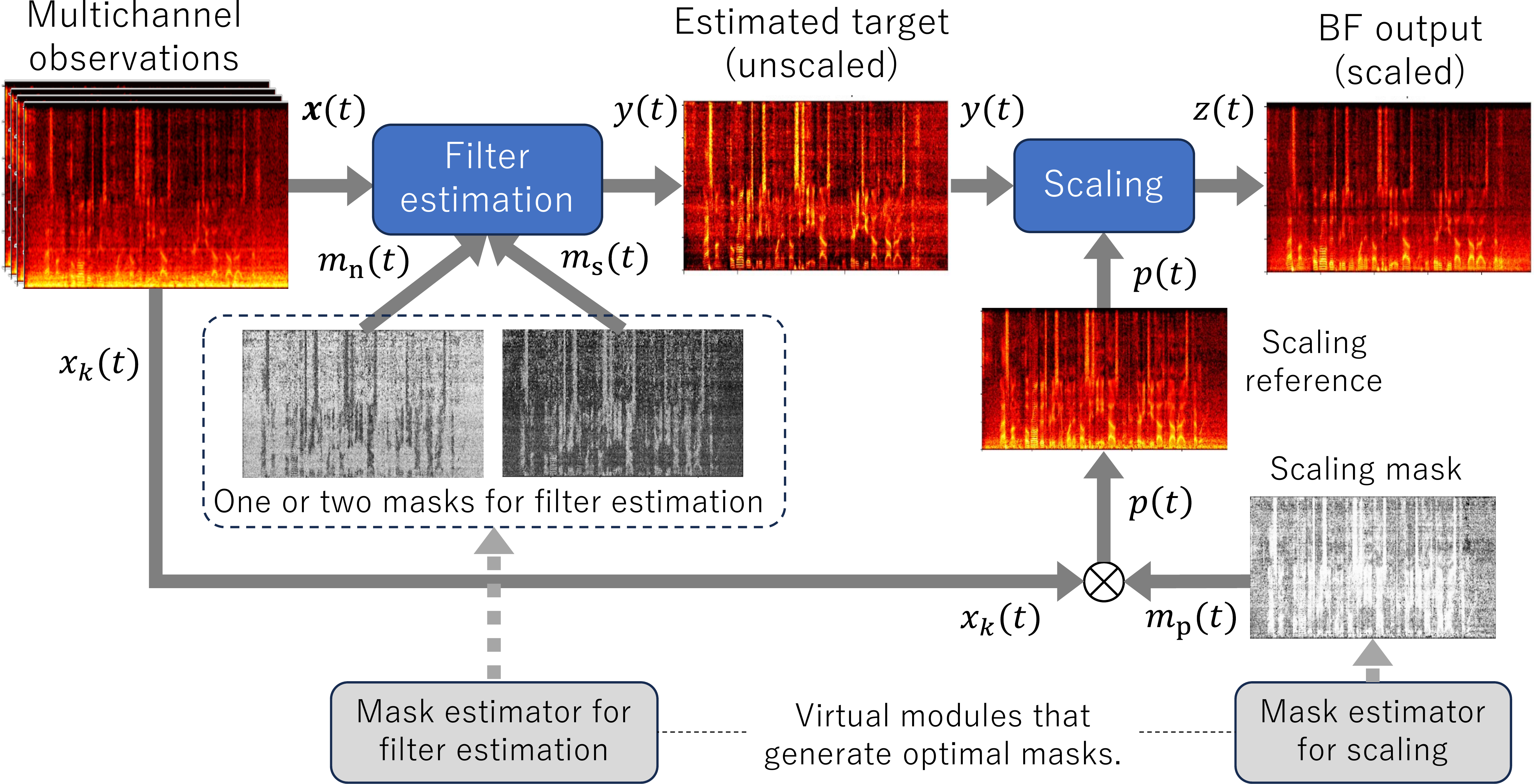}
  \caption{Unified framework of mask-based BFs (proposed); this consists of two mask-based processes: filter estimation and scaling. This study assumes that the optimal masks for both processes are generated with virtual estimators.}
  \label{fig:scheme}
\end{figure}

In this study, we propose a unified framework of the mask-based BFs that addresses the two issues mentioned in \refsec{sec:intro}. The framework comprises two processes: filter estimation and scaling as illustrated in \reffig{fig:scheme}. The former process estimates an extraction filter and applies it to the observations to generate the estimated target; the latter process adjusts both the magnitude and phase of the BF output using a scaling reference. A characteristic of the framework is that both processes are mask-based; the filter estimation process employs one or two masks, corresponding to $\ms(t)$, $\mn(t)$, or both, depending on the variation used. The scaling process adopts an alternative mask $m_{\rm p}(t)$, called a {\it scaling mask}, to generate the scaling reference. This study considers that these masks are generated with virtual modules that estimate the optimal masks for the given observation data.

The framework is explained in the subsequent subsections. In \refsubsec{subsec:filter-estimation}, we consider how the filter estimation process can cover all possible variations. In \refsubsec{subsec:scaling}, we propose a mask-based scaling process that can be combined with any BF variations. In \refsubsec{subsec:constraints}, we examine proper mask types for the processes.

\subsection{Filter estimation process covering all variations} \label{subsec:filter-estimation}

In this study, the filter estimation process covers all BF variations to explore their peak extraction performance. Thus, we modify existing BF formulas shown in Tables \ref{tab:formulas-mask-based-bfs} and \ref{tab:formulas-ica-based-tse} to fit this process.

\begin{table}[b]
  \centering
\caption{Classification rule for identifying BF variations; each variation mane consists of both prefix (type name) representing the operators used and suffix representing covariance matrices used.}
\label{tab:prefix-and-suffix}
\begin{tabular}{lll}
\toprule
       & Name & Meaning \\
\midrule\midrule
Prefix & MaxGEV & \underline{Max}imum eigenvector in \underline{g}eneralized \underline{e}igen\underline{v}alue decomposition \\
       & MinGEV & \underline{Min}imum eigenvector in \underline{g}eneralized \underline{e}igen\underline{v}alue decomposition \\
       & INV & Matrix \underline{inv}ersion \\
       & ISEV & Matrix \underline{i}nversion and \underline{s}tandard \underline{e}igen\underline{v}alue decomposition \\
\midrule
Suffix & NS & $\phin$ and $\phis$ \\
       & OS & $\phix$ and $\phis$ \\
       & NO & $\phin$ and $\phix$ \\
\bottomrule
\end{tabular}
\end{table}

We can eliminate any scalar factors that adjust the filter scale, given that this is estimated in the subsequent process. Moreover, we can remove $\vech$ by applying \refeqn{eqn:steering-vector}. Then, we rename the formulas using the rule shown in \reftable{tab:prefix-and-suffix} to classify BF variations. A variation name consists of the prefix (type name) and suffix that reflect the operators and covariance matrices included in the formula, respectively. Through this step, we found that Tables \ref{tab:formulas-mask-based-bfs} and \ref{tab:formulas-ica-based-tse} do not include variations corresponding to INV-NO and ISEV-NO. Thus, we also examine the two as novel BF variations.

\begin{table}[t]
  \centering
\caption{All possible variations of mask-based BFs and corresponding conventional methods; INV- and ISEV-NO BFs have not been employed.}
\label{tab:12variations}
\begin{tabular*}{\textwidth}{@{\extracolsep\fill}lllp{10em}}
\toprule
Variation name & Filter estimation & Masks used &  Corresponding conventional methods\\
\midrule\midrule
MaxGEV-NS & $\vecw=\gevmax{\phis}{\phin}$ &$\mn(t),\ms(t)$&
Max-SNR \\
MaxGEV-OS  & $\vecw=\gevmax{\phis}{\phix}$ &$\ms(t)$& Max-SOR \\
MaxGEV-NO  & $\vecw=\gevmax{\phix}{\phin}$ &$\mn(t)$& Max-ONR \\
\midrule
MinGEV-NS & $\vecw=\gevmin{\phin}{\phis}$ &$\mn(t),\ms(t)$& Min-NSR\\
MinGEV-OS  & $\vecw=\gevmin{\phix}{\phis}$ &$\ms(t)$& Min-OSR \\
MinGEV-NO  & $\vecw=\gevmin{\phin}{\phix}$ &$\mn(t)$& Min-NOR, SIBF \\
\midrule
INV-NS & $\vecw=\phin^{-1}\phis\vece_{k}$ &$\mn(t),\ms(t)$& Souden MVDR \\
INV-OS  & $\vecw=\phix^{-1}\phis\vece_{k}$ &$\ms(t)$& MMSE \\
INV-NO  & $\vecw=\phin^{-1}\phix\vece_{k}$ &$\mn(t)$& (Novel)\\
\midrule
ISEV-NS & $\vecw=\phin^{-1}\sevmax{\phis}$ &$\mn(t),\ms(t)$& MVDR, MLDR \\
ISEV-OS  & $\vecw=\phix^{-1}\sevmax{\phis}$ &$\ms(t)$& MPDR \\
ISEV-NO  & $\vecw=\phin^{-1}\sevmax{\phix}$ &$\mn(t)$& (Novel) \\
\bottomrule
\end{tabular*}
\end{table}

\begin{table}[t]
  \centering
\caption{Trivial optimal masks for INV-NS, OS, and NO BFs; note that these are complex-valued, so non-negative and more constrained masks cannot take these values.}
\label{tab:trivial-optimal-masks}
\begin{tabular}{ll}
\toprule
Variation name & Mask value\\
\midrule\midrule
INV-NS & $\dfrac{\ms(t)}{\mn(t)} = \dfrac{\htp{\vecx(t)}\vecw_{\rm ideal}}{\conj{x_{k}(t)}}$\\
\midrule
INV-OS  & $\ms(t) = \dfrac{\htp{\vecx(t)}\vecw_{\rm ideal}}{\conj{x_{k}(t)}}$ \ or \ $\ms(t) = \dfrac{\;\conj{s_{k}(t)}\;}{\;\conj{x_{k}(t)}\;}$\\
\midrule
INV-NO  & $\mn(t) = \dfrac{\conj{x_{k}(t)}}{\htp{\vecx(t)}\vecw_{\rm ideal}}$\\
\bottomrule
\end{tabular}
\end{table}

Consequently, we obtain 12 variations including two novel ones as shown in \reftable{tab:12variations}. We have several points to note about the variations. The MinGEV type is theoretically equivalent to the MaxGEV type, as mentioned in \refsubsec{subsec:max-snr}; thus, we only need to consider one. This study examines the MinGEV type to match the order of the covariance matrices in the formulas with other types. Therefore, the number of variations is practically nine. In contrast, the INV and ISEV types do not contain any equivalent pairs. Moreover, given that \refeqn{eqn:optimal-mask-def} indicates a different minimization problem for each variation, the optimal mask for a variation differs from that for the others except for the equivalent pairs.

Then, we discuss the trivial optimal masks that obtain the same filter as the ideal MMSE BF. Three variations belonging to the INV type contain the trivial optimal masks if the masks are not constrained, as discussed in Appendix \ref{sec:trivial-optimal-masks-for-inv-ns}. \reftable{tab:trivial-optimal-masks} shows the corresponding mask values, indicating that these are complex-valued. However, non-negative or more constrained masks cannot achieve these values. Furthermore, variations other than these three do not contain trivial optimal masks even if mask values are not constrained.

\subsection{Mask-based scaling process} \label{subsec:scaling}
This study employs a unified scaling method for all the BF variations. We define the scaling process as approximating the target by multiplying $y(t)$ by a scaling factor $\gamma$ in \refeqn{eqn:scaling-def}. This study considers that the process adjusts both the magnitude and the phase of $y(t)$. Therefore, $\gamma$ needs to be complex-valued.

Considering that the IS does not apply to realistic scenarios, as mentioned in \refsubsec{subsec:scaling-methods}, this study requires an alternative method that satisfies the following criteria: 1)~independent of the BF variation used, 2)~comparable with the IS, and 3)~free from the target. Therefore, we propose mask-based scaling formulated as follows:
\begin{eqnarray}
p(t) &=& \maskp(t) x_{k}(t),
        \label{eqn:scaling-reference}\\
\gamma &=& \argmin_{\gamma} \average{\left|p(t) - z(t)\right|^{2}}_{t},
             \label{eqn:scaling-factor-def} \\
        &=& \frac{\average{p(t) \conj{y(t)}}_{t}}{\average{\left|y(t)\right|^{2}}_{t}},
             \label{eqn:scaling-factor-value}
\end{eqnarray}
where $p(t)$ and $\maskp(t)$ denote a scaling reference and scaling mask, respectively. Note that this method is linear processing different from the post-masking that calculates $z(t)=\maskp(t)y(t)$~\cite{Erdogan2016-jc,Gannot2017-dt,Saric2022-dv}. The mask-based scaling method can be combined with any BFs and applied to realistic scenarios by providing $\maskp(t)$. Moreover, this includes both IS and MDP as particular cases: $p(t)=s_{k}(t)$ and $p(t)=x_{k}(t)$ in \refeqn{eqn:scaling-factor-value}, which means $\maskp(t)=s_{k}(t)/x_{k}(t)$ and $\maskp(t)=1$ in \refeqn{eqn:scaling-reference}, respectively.

Similar to the optimal mask for the filter estimation represented as \refeqn{eqn:optimal-mask-def}, we consider that the optimal scaling mask is the solution to the following minimization problem under a proper mask value constraint:
\begin{equation}
\maskset_{\rm p} = \argmin_{\maskset_{\rm p}} \average{\left|s_{k}(t) - z(t)\right|^{2}}_{t},
\label{eqn:optimal-scaling-mask-def}
\end{equation}
where $\maskset_{\rm p}$ denotes a set of $\maskp(t)$ over all frames.

The trivial optimal mask for scaling is discussed. If $\maskp(t)$ can take any complex value, $\maskp(t)=s_{k}(t)/x_{k}(t)$ is evidently optimal because this makes \refeqn{eqn:scaling-factor-value} identical to \refeqn{eqn:ideal-scaling}. However, non-negative or more constrained masks cannot contain this complex value. Thus, the optimal mask is not evident for these mask types. Moreover, considering that the scale of $\maskp(t)$ affects $\gamma$, a more constrained mask may degrade extraction performance due to inaccurate scaling. Therefore, we experimentally explore an appropriate mask type.

Additionally, the relationship between mask-based scaling and MMSE (or INV-OS) BF is considered. When both methods are combined, that is, \refeqn{eqn:bf-output} and \refeqn{eqn:mask-mmse} are applied to \refeqn{eqn:scaling-factor-value}, $\gamma$ is represented as
\begin{eqnarray}
\gamma &=& \frac{\htp{\vece_{k}}\average{\maskp(t)\vecx(t)\htp{\vecx(t)}}_{t}\phix^{-1}\phis\vece_{k}}%
     {\htp{\vece_{k}}\average{\conj{\ms(t)}\vecx(t)\htp{\vecx(t)}}_{t}\phix^{-1}\phis\vece_{k}}. \label{eqn:mmse-and-scaling}
\end{eqnarray}
The case $\maskp(t)=\conj{\ms(t)}$ for all $t$ results in $\gamma=1$. This fact indicates that mask-based MMSE BF involves the effect of mask-based scaling if $\maskp(t)=\conj{\ms(t)}$. Similarly, the ideal MMSE BF includes the effect of IS because both correspond to the case $\maskp(t)=\conj{\ms(t)}=s_{k}(t)/x_{k}(t)$.

\subsection{Proper mask types for the framework} \label{subsec:constraints}
A key issue in the unified framework is determining the appropriate mask type for $\ms(t)$, $\mn(t)$, and $\maskp(t)$. We address this issue from three perspectives, as shown in \reftable{tab:mask-constraints-this-study}.

\begin{table}[t]
  \centering
\caption{Mask types used in this study; ratio mask is examined for filter estimation, whereas non-negative, L1-MN, L2-MN, and ratio masks are compared for scaling, regardless of the original formulation.}
\label{tab:mask-constraints-this-study}
\begin{tabular*}{\textwidth}{@{\extracolsep\fill}lllp{0.135\textwidth}p{0.24\textwidth}}
\toprule
Process & Variation type & Mask & Required in formulation & Examined in this study \\
\midrule\midrule
Filter estimation & MaxGEV, MinGEV & $\ms(t)$ & Non-negative & Ratio\\
                  &                & $\mn(t)$ & Non-negative & Ratio\\
\cmidrule(l){2-5}
                  & INV            & $\ms(t)$ & Complex      & Ratio\\
                  &                & $\mn(t)$ & Complex & Ratio\\
\cmidrule(l){2-5}
                  & ISEV           & $\ms(t)$ & Complex      & Ratio\\
                  &                & $\mn(t)$ & Non-negative & Ratio\\
\midrule
Scaling           &             & $\maskp(t)$ & Complex & Non-negative, L1 MN, L2 MN, Ratio\\
\bottomrule
\end{tabular*}
\end{table}

1. Constraints in DNN training. Although this study does not include DNN training, it is important to consider this aspect because mask values are typically estimated using DNNs in real scenarios; for example, we can consider that the mask estimators in \reffig{fig:scheme} are properly trained DNNs that generate the masks from the observations. More constrained supervisory data can lead to more efficient training by integrating these constraints into the DNN structure, including the output-layer activation function~\cite{Jagtap2022-dv}. For example, when training with ratio masks, using a sigmoid function in the output layer can enhance training efficiency~\cite{Erdogan2016-jc,Zhang2017-nm}. Similarly, for non-negative data, incorporating an activation function that outputs non-negative values can improve training efficiency~\cite{Jagtap2022-dv,Mashrur2021-ft}. In summary, complex-valued masks are unnecessary if non-negative and more constrained masks can achieve the theoretical upper-bound performance, as no constraints can be applied to training with complex-valued masks.

2. Filter estimation. As discussed in Appendix \ref{sec:formulation-linear-tse}, different BFs require different constraints on the masks used for filter estimation. However, to compare all BF variations under unified conditions, the framework uses ratio masks for several reasons. For the MaxGEV and MinGEV types, masks must be non-negative because they derive from the max-SNR BF, as explained in Appendix \ref{subsec:max-snr}. Ratio masks are used because the range of mask values does not affect the eigenvectors in \refeqn{eqn:gev}. For the INV and ISEV types, at least $\ms(t)$ can take any complex value as described in \refsubsec{subsec:mmse} and \refsubsec{subsec:mvdr}. We standardize the masks for these types to ratio masks to maintain consistency with the MaxGEV and MinGEV types. The distinction between non-negative and more constrained masks mainly affects the scale of $\vecw$ which can be adjusted during scaling. Our focus is on whether the constraint on $\ms(t)$ impacts the peak extraction performance for INV and ISEV variations.

3. Scaling. The discussion in \refsubsec{subsec:scaling} suggests that the complex-valued mask need not be employed because this type simply obtains the trivial optimal mask; thus, more constrained mask types are required. Both \refeqn{eqn:scaling-reference} and \refeqn{eqn:scaling-factor-value} indicate that the value range of $\maskp(t)$ influences the BF output. Therefore, we need to examine the non-negative, L1-MN, and L2-MN, ratio masks.

\section{Experiments} \label{sec:experiments}

To explore the peak extraction performance for all variations described in \reftable{tab:12variations} and verify whether the mask-based scaling is comparable with IS, we conducted a series of experiments using the unified framework. Considering that the framework consists of two processes, filter estimation and scaling, experiments were conducted as follows:
\begin{enumerate}
    \item Exploring the relationship between iterations (counts of updating the masks) and extraction performance.
    \item Comparing all BF variations employing the IS.
    \item Comparing six scaling methods: mask-based scaling using non-negative, L1-MN, L2-MN, and ratio masks, as well as IS and MDP.
    \item Jointly optimizing each variation and the L1-MN-mask-based scaling.
\end{enumerate}
The setups for each experiment are shown in \reftable{tab:exp-setup}, and explained later.

In the subsequent subsections, we describe the dataset and common setups used in the experiments and demonstrate the experimental results in order.

\begin{table}[b]
\centering
\caption{Experimental setups (FE: filter estimation; 9 variations: variations other than MaxGEV type shown in \reftable{tab:12variations}; 12 variations: all variations shown in \reftable{tab:12variations}; Dev.: development set; 4 metrics: SDR, PESQ, STOI, and eSTOI; JO: joint optimization)}
\label{tab:exp-setup}
\begin{tabular*}{\textwidth}{@{\extracolsep\fill}clllllcc}
\toprule
\multicolumn{1}{c}{Section} & \multicolumn{1}{c}{Dataset} & \multicolumn{1}{c}{$g$} & \multicolumn{1}{c}{FE} & \multicolumn{1}{c}{Scaling} & 
\multicolumn{1}{c}{Metric} & \multicolumn{2}{c}{Iterations}\\
\cmidrule(r){7-8}
 & & & & & & \multicolumn{1}{c}{FE} & \multicolumn{1}{c}{Scaling} \\
\midrule\midrule
\refsubsec{subsec:exp-1}
  & Dev. & 1 & 9 variations & IS & SDR & 50--500 & - \\
\midrule
\refsubsec{subsec:compare-nine-variations}
  & Dev. & 1, 2, 4 & 12 variations & IS & SDR & 500\footnotemark[1] & - \\
\midrule
\refsubsec{subsec:compare-scaling-methods}
  & Dev. & 1, 2, 4 & Ideal MMSE & \multicolumn{1}{p{8em}}{Non-negative, L1-MN, L2-MN, Ratio, IS, MDP} & SDR & - & 500\\
\midrule
\refsubsec{subsec:compare-separate-and-joint}
  & Dev.    & 1, 2, 4 & 9 variations & L1-MN & SDR & \multicolumn{2}{c}{500\footnotemark[1] (JO)} \\
  & Test    & 1        & 9 variations& L1-MN & 4 metrics       & \multicolumn{2}{c}{500\footnotemark[1] (JO)}\\
\bottomrule
\end{tabular*}
\footnotetext[1]{Exceptionally, ISEV-OS used 1000 iterations because of slower convergence.}
\end{table}

\subsection{Dataset and common setups} \label{subsec:dataset}

\begin{figure}[t]
  \centering
  \includegraphics[width=0.8\linewidth]{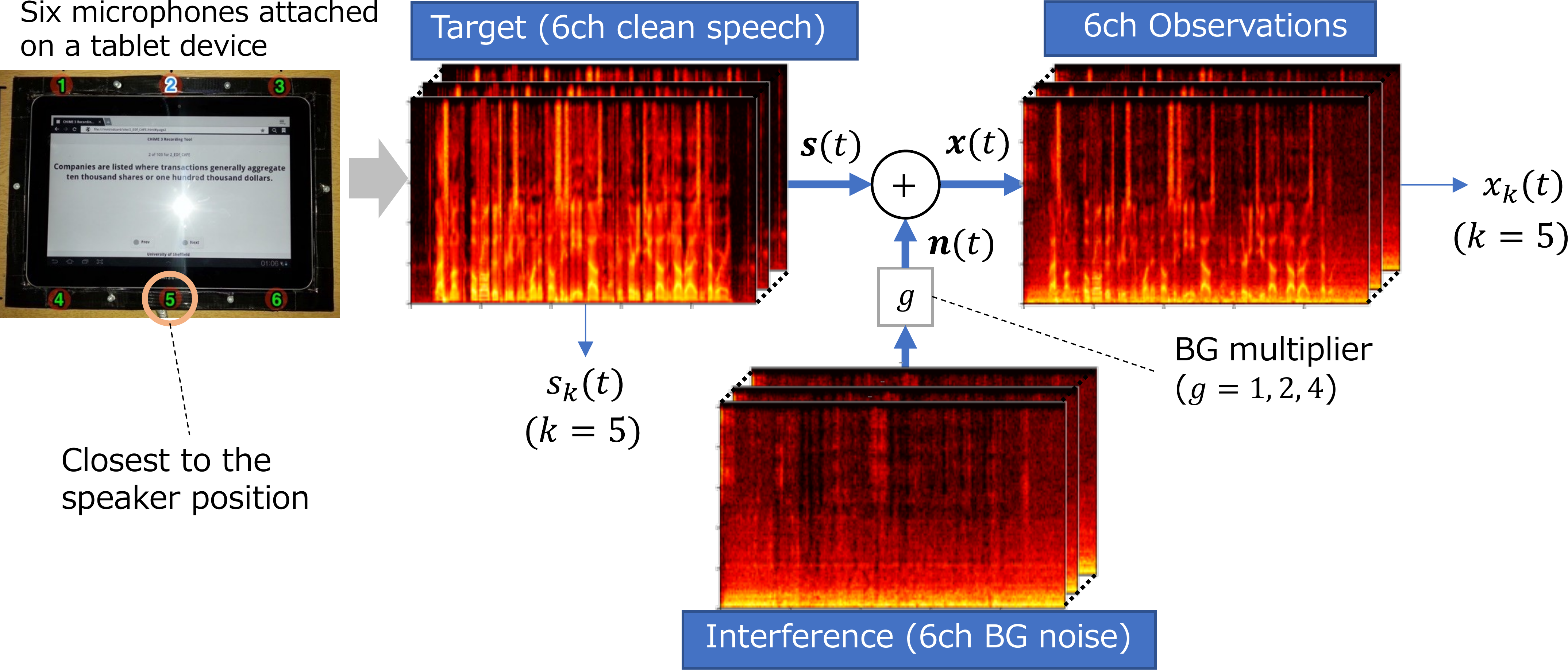}
  \caption{Process of generating observation data for three noisy scenarios; the multiplier $g$ was applied to the background (BG) noises before being mixed with the clean speeches to generate different noisy scenarios shown in \reftable{tab:snr-dev-set}.}
  \label{fig:mixing}
\end{figure}

We employed both the development and test sets included in the CHiME-4 simulated dataset~\cite{Vincent2017-px}. The same data were included in the CHiME-3 dataset~\cite{Barker2017-tr}. The development set contained 410 utterances from four speakers (1640 utterances in total) and four background (BG) noises. The sound data of this dataset was recorded at 16~kHz by six microphones attached to a tablet device. The speaker-tablet distance was typically around 40 cm~\cite{Barker2017-tr}. We generated the TF domain signals using short-time Fourier transform with window and shift lengths of 1024 and 256, respectively. To represent multiple scenarios in different SNRs, we artificially mixed the utterances and one of the background noises, applying three multipliers, $g=1.0$, 2.0, and 4.0, to the BG noise as shown in \reffig{fig:mixing}. We refer to these values as {\it BG multipliers}. Each scenario comprises 1640 utterances and its SNR score is indicated in \reftable{tab:snr-dev-set}. These scenarios were used for all experiments. Experiment 4 also used the CHiME-4 test set, comprising 330 utterances from four speakers and four BG noises (1320 utterances in total).

\begin{table}[t]
  \centering
\caption{SNR [dB] for each scenario; the development set consisted of three scenarios, whereas the test set contained a single one.}
\label{tab:snr-dev-set}
\begin{tabular*}{0.6\textwidth}{@{\extracolsep\fill}ccccc}
\toprule
& \multicolumn{3}{c}{Development} & Test \\
\cmidrule(r){2-4}\cmidrule(l){5-5}
BG Multiplier $g$ & 1.0 & 2.0 & 4.0 & 1.0\\
\midrule
SDR [dB]       & 5.79    & -0.21   & -6.12 & 7.54\\
\bottomrule
\end{tabular*}
\end{table}

All experiments used the SDR~\cite{Vincent2006-pg} as an evaluation metric, calculated as follows:
\begin{equation}
    {\rm SDR~[dB]} = 10 \log_{10} \left(
    \frac{\average{|S_{k}(t)|^2}_{t}}{\average{|S_{k}(t)-Z(t)|^2}_{t}}
    \right), \label{eqn:sdr}
\end{equation}
where $S_{k}(t)$ and $Z(t)$ denote the waveforms corresponding to $s_{k}(t)$ and $z(t)$, respectively.
Experiment 4 also used the narrowband perceptual evaluation of speech quality (PESQ)~\cite{Beerends2013-wx}, short-time objective intelligibility measure (STOI)~\cite{Taal2011-wo}, and extended STOI (eSTOI)~\cite{Jensen2016-un}. Basically, these four metrics show higher scores as the BF output approaches the target.

Considering that Microphone \#5 was the closest to the speaker position as illustrated in \reffig{fig:mixing}, $k$ was set to 5 as the reference microphone index. That is, $k=5$ was used in all formulas in \reftable{tab:12variations}, and in \refeqn{eqn:optimal-mask-def}, \refeqn{eqn:ideal-mmse}, \refeqn{eqn:ideal-scaling}, and \refeqn{eqn:scaling-reference}. Similarly, $S_{5}(t)$ was used as the reference signal for calculating the SDR, PESQ, STOI, and eSTOI scores.

All the systems employed in the experiments were implemented in PyTorch~\cite{Paszke2019-dm}, which supports the backpropagation of matrix operations in the complex number domain.

\subsection{Experiment 1: Exploring the relationship between iteration count and extraction performance} \label{subsec:exp-1}

\begin{figure}[b]
  \centering
  \includegraphics[width=0.7\linewidth]{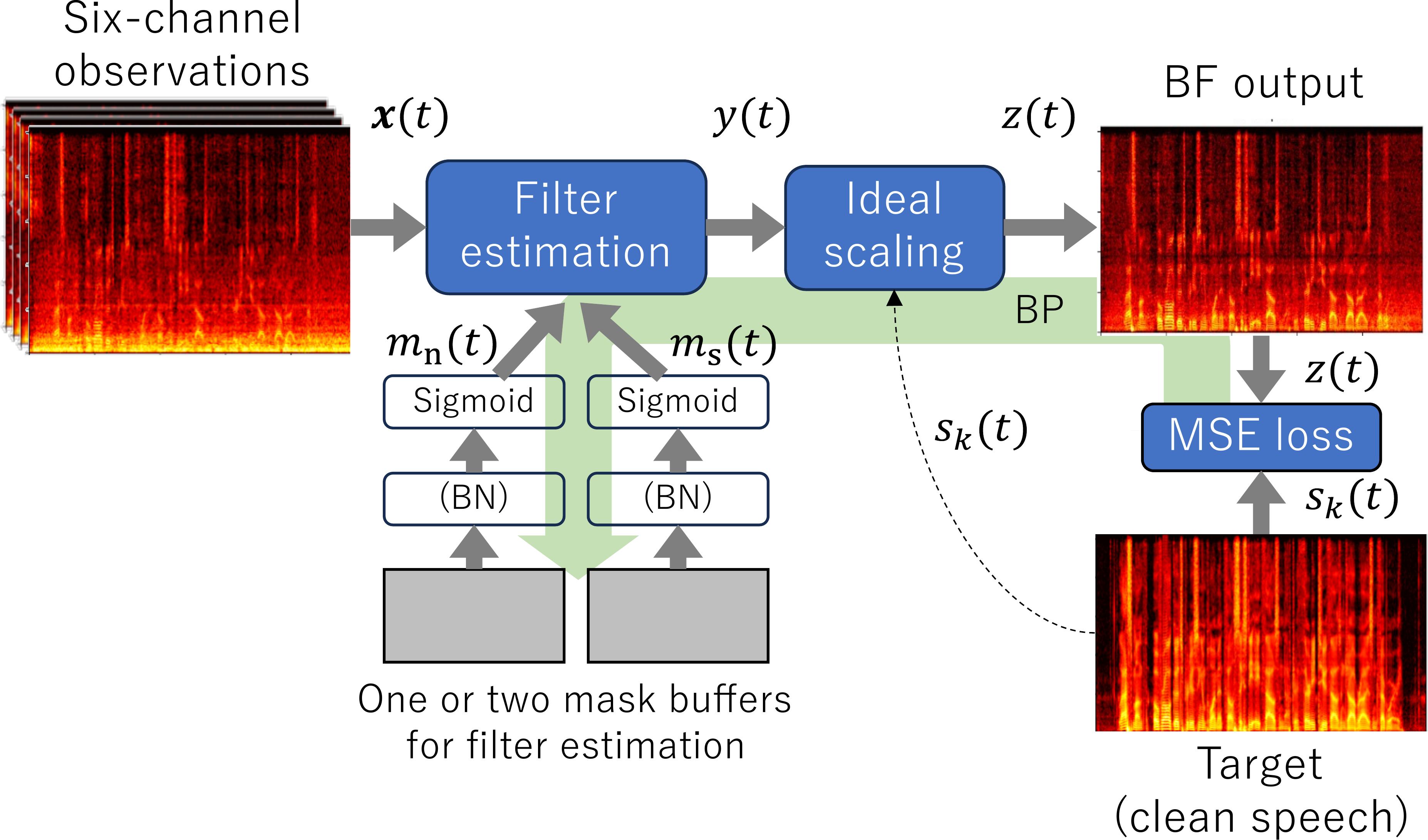}
  \caption{System used in Experiments 1 and 2; for filter estimation, one or two mask buffers were provided depending on the variation used. Buffered mask values were iteratively updated by backpropagation (BP) to minimize the MSE loss. The effect of batch normalization (BN) was examined.}
  \label{fig:exp1and2}
\end{figure}

First, we verified the following aspects using the setups listed in the first row of \reftable{tab:exp-setup}:
\begin{enumerate}
    \item How many iterations are sufficient for convergence?
    \item Can the batch normalization (BN) layer accelerate convergence?    
\end{enumerate}
The experimental system is illustrated in \reffig{fig:exp1and2}. In filter estimation, we examined nine variations other than the MaxGEV-NS, OS, and NO BFs out of the 12 shown in \reftable{tab:12variations}, considering the equivalence between the MaxGEV and MinGEV types. The scaling process was fixed to IS. One or two mask buffers were prepared depending on the variation used. We applied the sigmoid function to the buffered values to constrain the mask type to the ratio mask. The values were iteratively updated using backpropagation (BP) to minimize the MSE between the BF output $z(t)$ and target $s_{k}(t)$ ($k=5$) in \refeqn{eqn:optimal-mask-def}. The optimal mask for each variation was obtained on an utterance-by-utterance basis.

Since the sigmoid function can be interpreted as an activation function in the output layer~\cite{Jagtap2022-dv}, we inserted a BN layer~\cite{Ioffe2015-gq} before applying the function to achieve faster convergence. This layer treated each frequency bin as a BN channel. Unlike DNN training, the BN parameters were determined for each utterance. We compared both cases in which the BN was enabled and disabled.

\begin{figure}[t]
  \centering
  \includegraphics[width=1.0\linewidth]{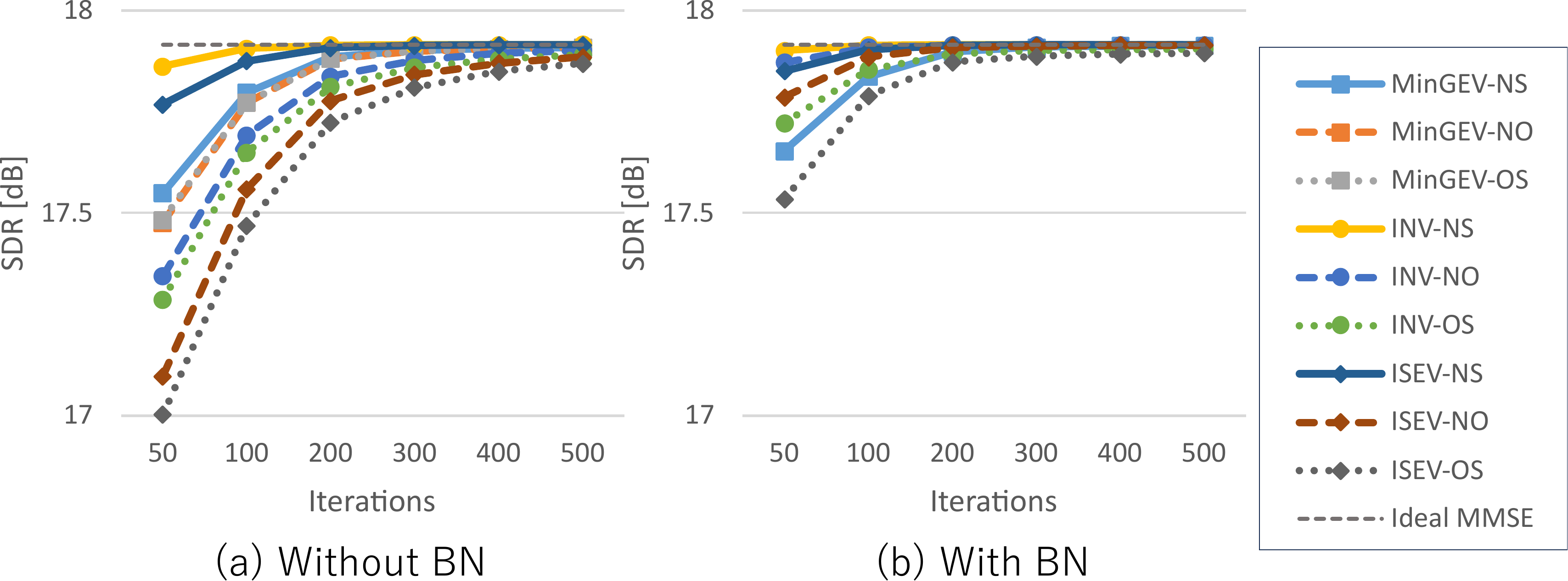}
  \caption{Relationship between SDR scores [dB] and iterations; batch normalization (BN) was omitted in (a) and enabled in (b). BN accelerated convergence, although it caused errors in MinGEV-NO and OS BFs.}
  \label{fig:rel-iteration}
\end{figure}

The relationships between each iteration count (ranging from 50 to 500) and the SDR score [dB] are plotted in \reffig{fig:rel-iteration}. In Part (a), the BN layer was omitted, whereas in Part (b), it was enabled. To illustrate the theoretical upper-bound performance, the score of the ideal MMSE BF (17.92 dB) is also plotted as a dashed line. The MinGEV-NO and OS BFs were excluded in Part (b) because they caused errors during the execution of the GEV.

Comparing Parts (a) and (b) suggests that the BN layer can accelerate convergence. Therefore, we adopted 500 iterations with the BN except for the following three BFs; for MinGEV-NO and OS BFs, the BN layer was omitted because of the aforementioned error; for the ISEV-OS BF, 1000 iterations were adopted considering that its convergence was the slowest even enabling the BN.

\subsection{Experiment 2: Comparing all BF variations} \label{subsec:compare-nine-variations}
 
Next, we compared all the variations of the unified framework using the same system as Experiment 1 and the setups in the second row of \reftable{tab:exp-setup}. Unlike Experiment 1, all 12 variations shown in \reftable{tab:12variations} were evaluated using three noisy scenarios of the development set shown in \reftable{tab:snr-dev-set}. The ideal MMSE BF represented as \refeqn{eqn:ideal-mmse} was also evaluated to determine the theoretical upper-bound performance of the BFs.

\begin{table}[t]
  \centering
\caption{SDR scores [dB] of all BF variations with IS in Experiment 2 (BN: batch normalization); all the variations practically achieved the same scores comparable with the upper-bound obtained with the ideal MMSE BF.}
\label{tab:exp2}
\begin{tabular*}{0.76\textwidth}{@{\extracolsep\fill}lcrrrr}
\toprule
Variation name  & BN & Iterations & \multicolumn{1}{c}{$g=1.0$} & \multicolumn{1}{c}{$g=2.0$} & \multicolumn{1}{c}{$g=4.0$} \\
\midrule\midrule
MaxGEV-NS & $\checkmark$ & 500 & 17.91 & {\bf 12.64} & {\bf 7.74} \\
MaxGEV-NO &  & 500 & 17.91 & {\bf 12.64} & {\bf 7.74} \\
MaxGEV-OS &  & 500 & 17.91 & {\bf 12.64} & {\bf 7.74} \\
\midrule
MinGEV-NS & $\checkmark$ & 500 & 17.91 & {\bf 12.64} & {\bf 7.74} \\
MinGEV-NO &  & 500 & 17.91 & {\bf 12.64} & {\bf 7.74} \\
MinGEV-OS &  & 500 & 17.91 & {\bf 12.64} & {\bf 7.74} \\
\midrule
INV-NS & $\checkmark$ & 500 & {\bf 17.92} & {\bf 12.64} & {\bf 7.74} \\
INV-NO & $\checkmark$ & 500 & {\bf 17.92} & {\bf 12.64} & {\bf 7.74} \\
INV-OS & $\checkmark$ & 500 & 17.91 & 12.63 & 7.73 \\
\midrule
ISEV-NS & $\checkmark$ & 500 & {\bf 17.92} & {\bf 12.64} & {\bf 7.74} \\
ISEV-NO & $\checkmark$ & 500 & {\bf 17.92} & {\bf 12.64} & {\bf 7.74} \\
ISEV-OS & $\checkmark$ & 1000 & 17.90 & 12.62 & 7.72 \\
\midrule\midrule
Ideal MMSE & n/a & n/a & {\bf 17.92} & {\bf 12.64} & {\bf 7.74} \\
\bottomrule
\end{tabular*}
\end{table}

\reftable{tab:exp2} shows SDR scores of all variations and the ideal MMSE BF for the three scenarios. Given that the maximum SDR difference in this table is only 0.02 dB, we can regard that all variations achieved the same extraction performance comparable to the upper bound. Remarkably, both INV-NO and ISEV-NO BFs achieved the same performance even though not employed as BFs.

We also confirmed that MaxGEV-NS, NO, and OS BFs achieved the same performance as MinGEV-NS, NO, and OS BFs, respectively, because of the theoretical equivalence mentioned in \refsubsec{subsec:filter-estimation} and \refsubsec{subsec:max-snr}. Therefore, we did not examine the MaxGEV type in subsequent experiments.

\subsection{Experiment 3: Comparing the scaling methods} \label{subsec:compare-scaling-methods}

\begin{figure}[t]
  \centering
  \includegraphics[width=0.7\linewidth]{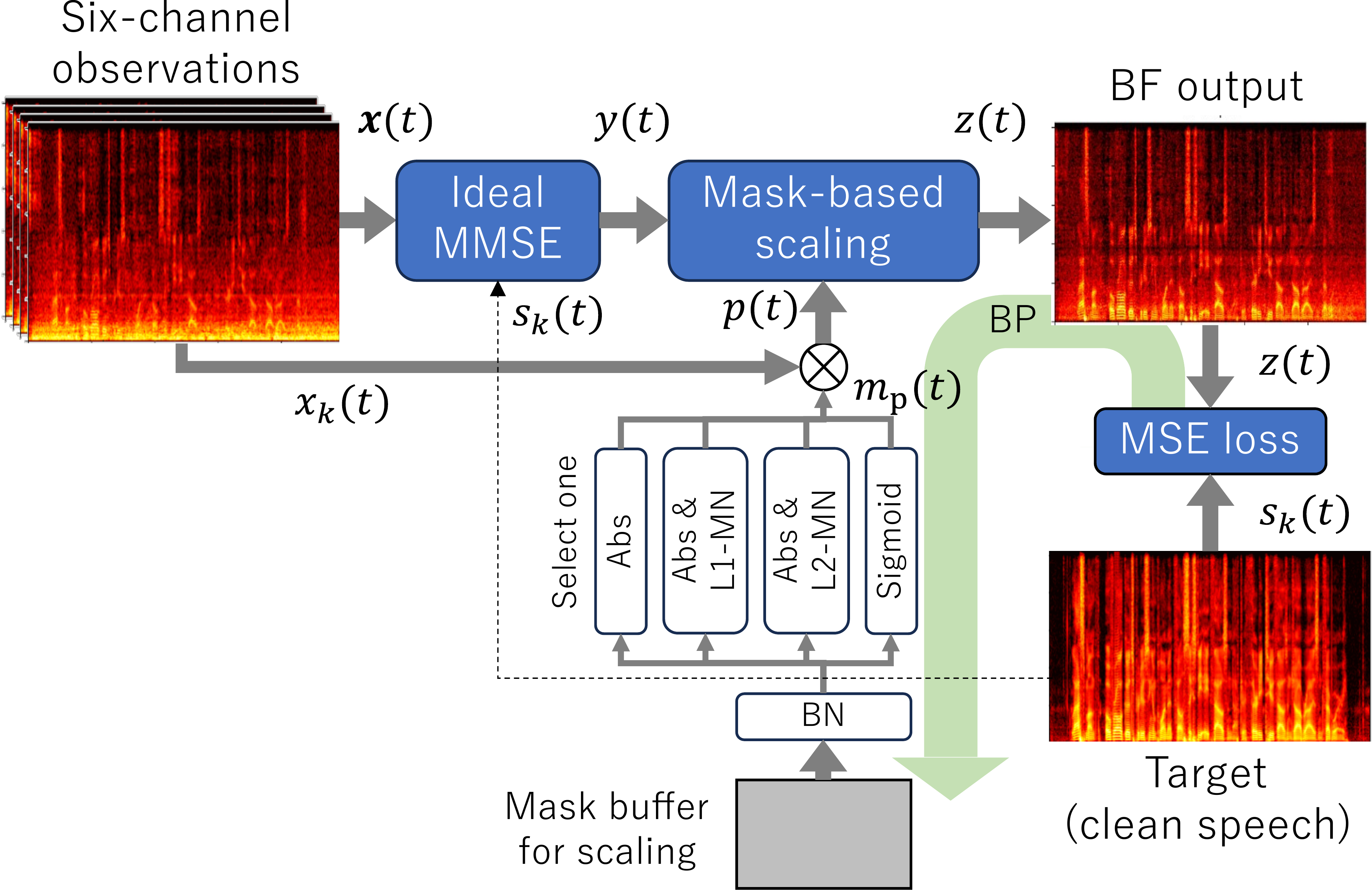}
  \caption{System used in Experiment 3 to evaluate the mask-based scaling (Abs: absolute function); modules labeled as `Abs,' `Abs\&L1-MN,' `Abs\&L2-MN,' and `Sigmoid' indicate constraints for the non-negative, L1-MN, L2-MN, and ratio masks, respectively.}
  \label{fig:exp3}
\end{figure}

Next, we compared the following six scaling methods, using the setups in the third row of \reftable{tab:exp-setup} and the system illustrated in \reffig{fig:exp3}:
\begin{itemize}
    \item Four setups of the mask-based scaling using non-negative, L1-MN, L2-MN, and ratio masks as $\maskp(t)$ in \refeqn{eqn:scaling-reference}
    \item IS for evaluating the upper-bound scaling performance
    \item MDP as a conventional method that can adjust both the magnitude and phase of the BF output.
\end{itemize}
The filter estimation process was fixed to the ideal MMSE BF represented as \refeqn{eqn:ideal-mmse}. The mask-based scaling process required a single mask buffer. BN was applied to the buffered values to accelerate their convergence. To generate the above four masks, we exclusively applied the following operations: 1) absolute function (Abs), 2) Abs and L1 mean normalization represented in \refeqn{eqn:l1-mean-normalized}, 3) Abs and L2 mean normalization represented in \refeqn{eqn:l2-mean-normalized}, and 4) sigmoid function. The optimal scaling mask was iteratively obtained by minimizing the MSE represented in \refeqn{eqn:optimal-scaling-mask-def}. For sufficient convergence, 500 iterations were adopted.

\reftable{tab:exp3} presents the SDR score for each scaling method in the three scenarios. The scores for the IS were identical to those of the ideal MMSE BF shown in Experiment 2 because the ideal MMSE BF inherently includes the effect of the IS, as mentioned in \refsubsec{subsec:scaling}. The scaling methods using non-negative, L1-MN, and L2-MN masks achieved the same scores as the IS, whereas the method using a ratio mask produced scores comparable to the non-negative mask or slightly lower. In contrast, the MDP method showed a larger degradation in performance compared with the others.

\begin{table}[t]
  \centering
\caption{SDR scores [dB] on comparing the scaling methods in Experiment 3; mask-based methods using the non-negative, L1-MN, and L2-MN masks achieved the same scores as the ideal scaling (IS), whereas one using the ratio mask slightly degraded. The minimal distortion principle (MDP) method degraded more largely.}
\label{tab:exp3}
\begin{tabular}{llrrr}
\toprule
Scaling method & Mask type & \multicolumn{1}{c}{$g=1.0$} & \multicolumn{1}{c}{$g=2.0$} & \multicolumn{1}{c}{$g=4.0$} \\
\midrule
Mask-based & Non-negative (Abs)& {\bf 17.92} & {\bf 12.64} & {\bf 7.74} \\
           & L1-MN & {\bf 17.92} & {\bf 12.64} & {\bf 7.74} \\
           & L2-MN & {\bf 17.92} & {\bf 12.64} & {\bf 7.74} \\
           & Ratio (Sigmoid) & 17.88       & 12.62       & 7.73 \\
\midrule
IS         & Complex ($m_{\rm p}(t)=s_{k}(t)/x_{k}(t)$)      & {\bf 17.92} & {\bf 12.64} & {\bf 7.74} \\
MDP (conventional) & $m_{\rm p}(t)=1$ for all $t$ & 17.23       & 11.33       & 5.38 \\
\bottomrule
\end{tabular}
\end{table}

In subsequent experiments, we adopted the L1-MN mask because 1) the method using this mask can achieve the same scores as the IS, 2) this mask is more constrained than the non-negative mask, and 3) the constraint for this mask represented as \refeqn{eqn:l1-mean-normalized} is simpler than that for the L2-MN mask, represented as \refeqn{eqn:l2-mean-normalized}.

\subsection{Experiment 4: Joint optimization} \label{subsec:compare-separate-and-joint}
\begin{figure}[t]
  \centering
  \includegraphics[width=0.7\linewidth]{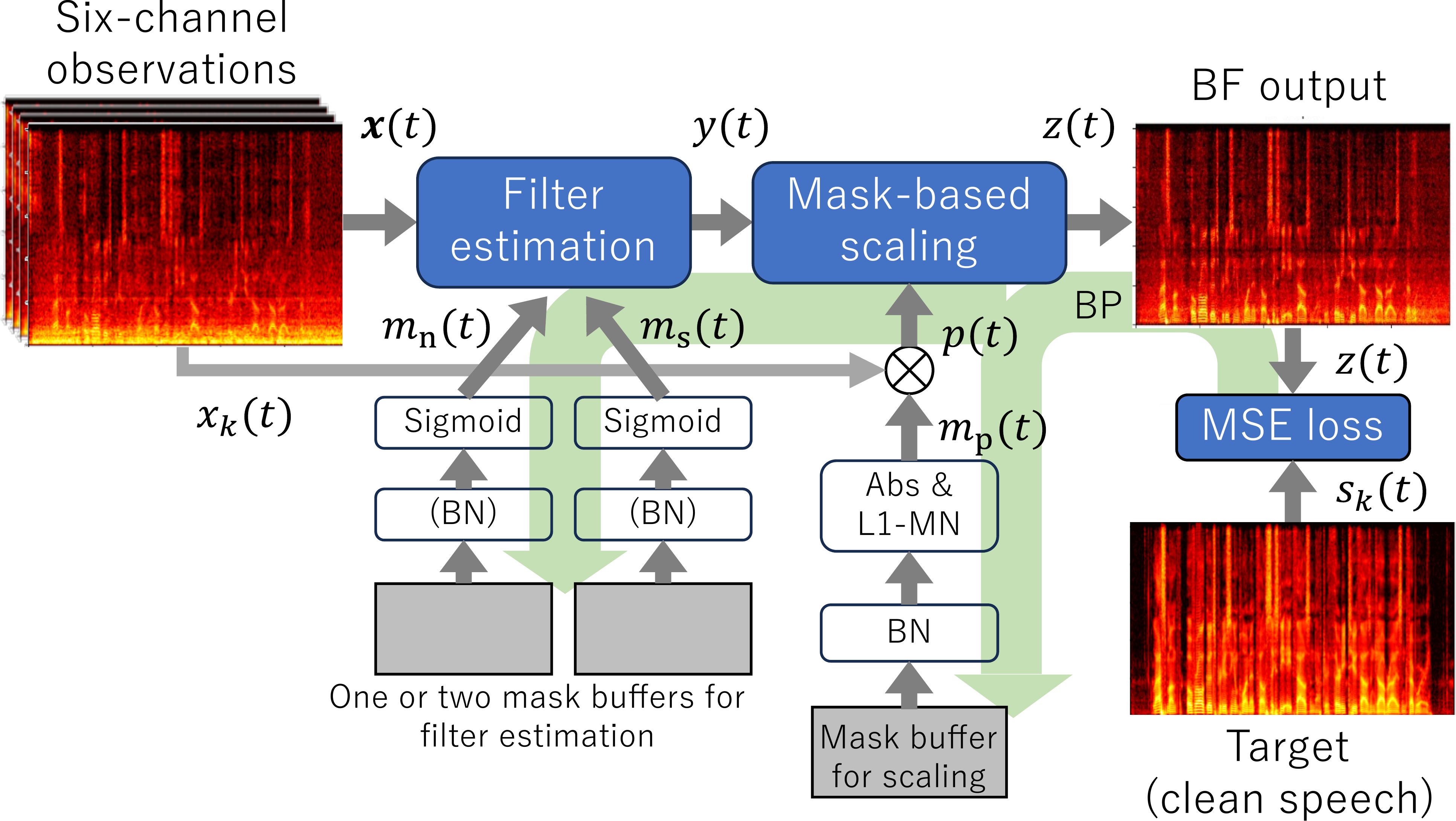}
  \caption{System used in Experiment 4; masks for both filter estimation and scaling were simultaneously optimized; nine variations were compared in the filter estimation, and the L1-MN mask was used in scaling.}
  \label{fig:exp4-2}
\end{figure}

Next, we obtained optimal masks for both the filter estimation and scaling processes using the fourth row of \reftable{tab:exp-setup}. \reffig{fig:exp4-2} illustrates the system used in this experiment. The filter estimation process was identical to that used in Experiment 2, although only nine variations were compared. The scaling process was the same as in Experiment 3, although only the L1-MN mask was used as the scaling mask.

A set of optimal masks for both processes was obtained as the solution to the following minimization problem:
\begin{equation}
\maskset_{\rm filt}, \maskset_{\rm p} = \argmin_{\maskset_{\rm filt}, \maskset_{\rm p}} \average{\left|s_{k}(t) - z(t)\right|^{2}}_{t}.
\label{eqn:joint-optimization}
\end{equation}
We employed 500 iterations with BN except for the cases of MinGEV-NO, MinGEV-OS, and ISEV-OS BFs, similar to Experiment 2.

\begin{table}[t]
  \centering
\caption{SDR scores [dB] of the joint optimization in Experiment 4; similar to Experiment 2, all the variations were comparable with the ideal MMSE BF, even using the mask-based scaling.}
\label{tab:exp4-dev}
\begin{tabular*}{0.75\textwidth}{@{\extracolsep\fill}lcrrr}
\toprule
Variation name & Equivalent to& $g=1.0$ & $g=2.0$ & $g=4.0$ \\
\midrule\midrule
MinGEV-NS & MaxGEV-NS & 17.91 & {\bf 12.64} & 7.74 \\
MinGEV-NO & MaxGEV-NO & 17.91 & {\bf 12.64} & 7.74 \\
MinGEV-OS & MaxGEV-OS & 17.91 & {\bf 12.64} & 7.74 \\
\midrule
INV-NS & - & {\bf 17.92} & {\bf 12.64} & {\bf 7.75} \\
INV-NO & - & 17.91 & 12.63 & 7.73 \\
INV-OS & - & 17.90 & 12.62 & 7.72 \\
\midrule
ISEV-NS & - & {\bf 17.92} & {\bf 12.64} & {\bf 7.75} \\
ISEV-NO & - & 17.91 & 12.63 & 7.73 \\
ISEV-OS & - & 17.91 & 12.62 & 7.72 \\
\midrule\midrule
Ideal MMSE & - & {\bf 17.92} & {\bf 12.64} & 7.74 \\
\bottomrule
\end{tabular*}
\end{table}

\reftable{tab:exp4-dev} presents the SDR scores for each variation in the three scenarios. Due to the equivalence mentioned in \refsubsec{subsec:compare-nine-variations}, we consider that the scores of MaxGEV-NS, NO, and OS BFs are the same as those of MinGEV-NS, NO, and OS, respectively. 
Although INV-NS and ISEV-NS BFs demonstrated a slightly larger score than the ideal MMSE in the $g=4.0$ scenario, we attribute this to an error caused by calculating the SDR in the time domain as represented in \refeqn{eqn:sdr}. Therefore, similar to Experiment 2, we can regard all the variations as achieving the theoretical upper-bound performance obtained with the ideal MMSE BF.

\begin{figure}[t]
  \centering
  \includegraphics[width=1.0\linewidth]{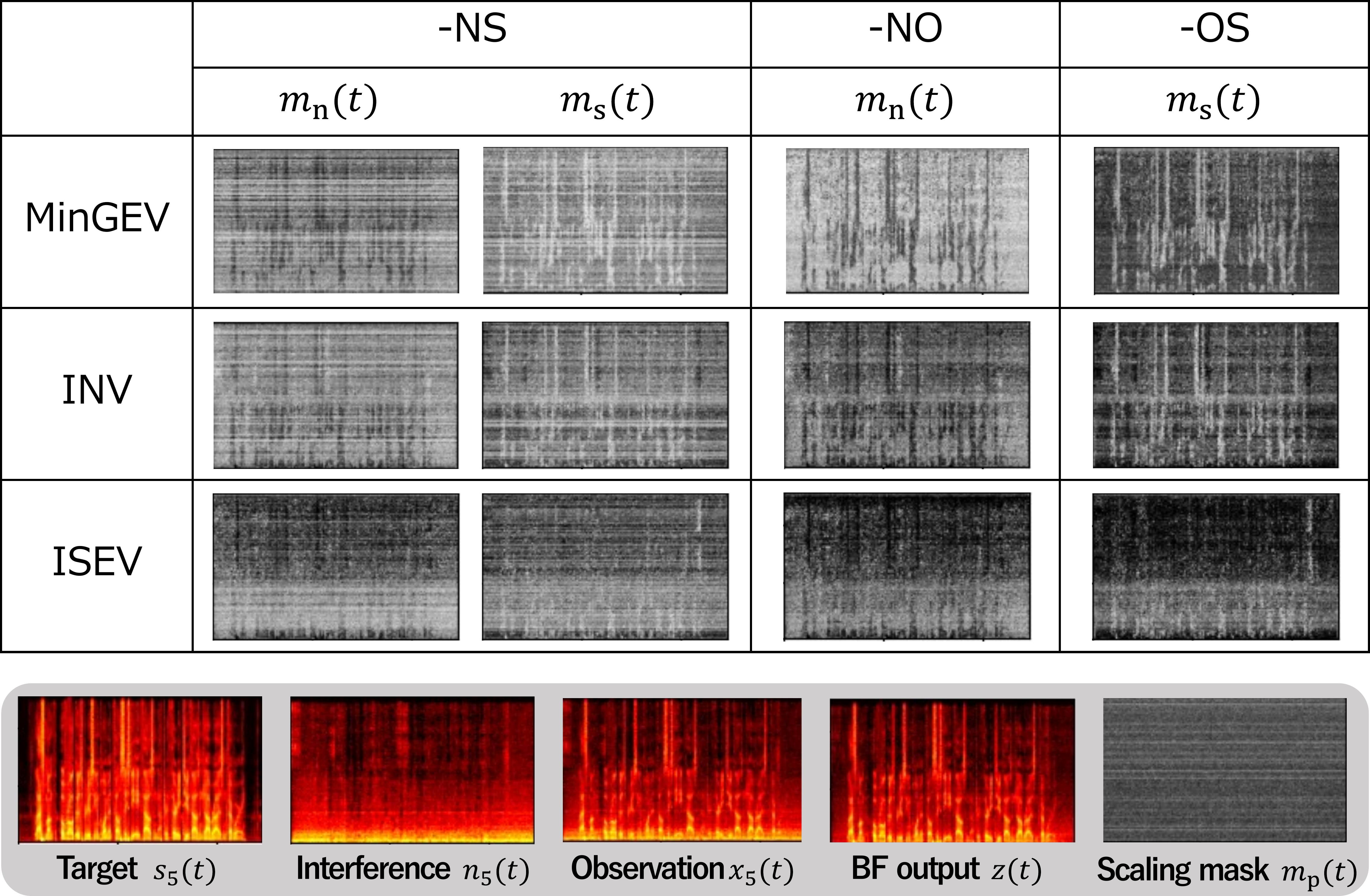}
  \caption{Optimal masks for nine BF variations obtained in Experiment 4 ($\ms(t)$: mask for the target, $\mn(t)$: mask for the interferences); the target, interference, observation, BF output, and optimal scaling mask are also displayed.}
  \label{fig:optimal-masks}
\end{figure}

\reffig{fig:optimal-masks} illustrates the optimal masks for the nine BF variations obtained in this experiment, along with the target $s_{5}(t)$, interferences (background noise) $n_{5}(t)$, observation $x_{5}(t)$, BF output $z(t)$, and optimal scaling mask $\maskp(t)$. The target is included in the development set and labeled {\it M04\_050C0101}. The interferences were recorded on a bus. The observation is a mixture of the target and interferences with $g=1$ in \reffig{fig:mixing}. Given that all the variations generated practically identical BF outputs and optimal scaling masks, \reffig{fig:optimal-masks} shows those obtained with the MinGEV-NS BF as representatives.

The optimal masks for filter estimation are ratio masks, and the mask values 0 and 1 are plotted in black and white, respectively. These appear to be different for each variation despite achieving the same performance. We discuss the results in \refsubsec{subsubsec:mask-unique}. Meanwhile, the optimal scaling mask is an L1-MN mask, with higher mask values plotted in brighter colors. As mentioned in \refsubsec{subsec:scaling}, this mask is the solution to the problem represented in \refeqn{eqn:optimal-scaling-mask-def}, independent of the BF variation employed. Consequently, the optimal scaling mask is practically identical for all BF variations.

\begin{table}[t]
  \centering
\caption{SDR [dB], PESQ, STOI [\%], and eSTOI [\%] scores using the CHiME-4 test set in Experiment 4; all the BF variations practically achieved the same extraction performance as the ideal MMSE in the four metrics.}
\label{tab:exp4-test}
\begin{tabular}{lcrrrr} 
\toprule
Variation name & Equivalent to
 & \multicolumn{1}{c}{SDR [dB]} & \multicolumn{1}{c}{PESQ} & \multicolumn{1}{c}{STOI [\%]} & \multicolumn{1}{c}{eSTOI [\%]} \\
\midrule\midrule
MinGEV-NS & MaxGEV-NS & 19.42 & {\bf 2.77} & {\bf 97.03} & 90.48 \\
MinGEV-NO & MaxGEV-NO & 19.43 & {\bf 2.77} & {\bf 97.03} & 90.48 \\
MinGEV-OS & MaxGEV-OS & 19.43 & {\bf 2.77} & {\bf 97.03} & 90.48 \\
\midrule
INV-NS & - & {\bf 19.44} & {\bf 2.77} & {\bf 97.03} & 90.48 \\
INV-NO & - & {\bf 19.44} & {\bf 2.77} & {\bf 97.03} & 90.48 \\
INV-OS & - & 19.43 & {\bf 2.77} & {\bf 97.03} & 90.48 \\
\midrule
ISEV-NS & - & {\bf 19.44} & {\bf 2.77} & {\bf 97.03} & 90.48 \\
ISEV-NO & - & 19.43 & {\bf 2.77} & 97.02 & 90.47 \\
ISEV-OS & - & 19.43 & {\bf 2.77} & {\bf 97.03} & 90.48 \\
\midrule\midrule
Microphone \#5 &-& 7.54 & 2.18 & 87.03 & 68.32 \\
Ideal MMSE &-& {\bf 19.44} & {\bf 2.77} & {\bf 97.03} & {\bf 90.49} \\
\bottomrule
\end{tabular}
\end{table}

Finally, we evaluated the same system on the CHiME-4 test set to measure four metrics: SDR, PESQ, STOI, and eSTOI. The results are shown in \reftable{tab:exp4-test}. Given that the maximum difference was just 0.02 points for all the metrics, we can consider that all the variations achieved the theoretical upper-bound performance in the test set and across the four metrics. The significance of these results is discussed in \refsubsec{subsec:comparison-using-chime}.

\section{Discussion} \label{sec:discussion}
The experimental results suggest the following aspects:
\begin{enumerate}
    \item All variations of the mask-based BFs using one or two ratio masks can achieve the theoretical upper-bound performance obtained with the ideal MMSE BF.
    \item The scaling process using an L1-MN or L2-MN mask can function as the IS.
    \item Jointly optimizing the masks can also achieve the upper-bound performance. This trend is verified in the SDR, PESQ, STOI, and eSTOI scores using the CHiME-4 test set.
\end{enumerate}
This section discusses these aspects in the subsequent subsections. Additionally, in \refsubsec{subsec:specific-discussion-on-several-variations}, we explore why several variations that have not traditionally been employed as BFs can still effectively extract the target.

\subsection{Why can all variations achieve the theoretical upper-bound performance?}

The experimental results suggest that all 12 variations can achieve the theoretical upper-bound performance even when the mask type is constrained to a ratio mask. We first explain that these results do not contradict studies that compared multiple BFs and reported different ones as the best~\cite{Heymann2016-sy,Boeddeker2018-ww,Wang2018-tl,Shimada2019-xj,Heymann2018-mf}. Then, we discuss the reason for achieving the same performance, using the concept of the practical parameter count and performance saturation point. Finally, we consider the uniqueness of the optimal mask.

\subsubsection{Explanation of non-contradiction with previous studies}

\begin{figure}[t]
  \centering
  \includegraphics[width=0.5\linewidth]{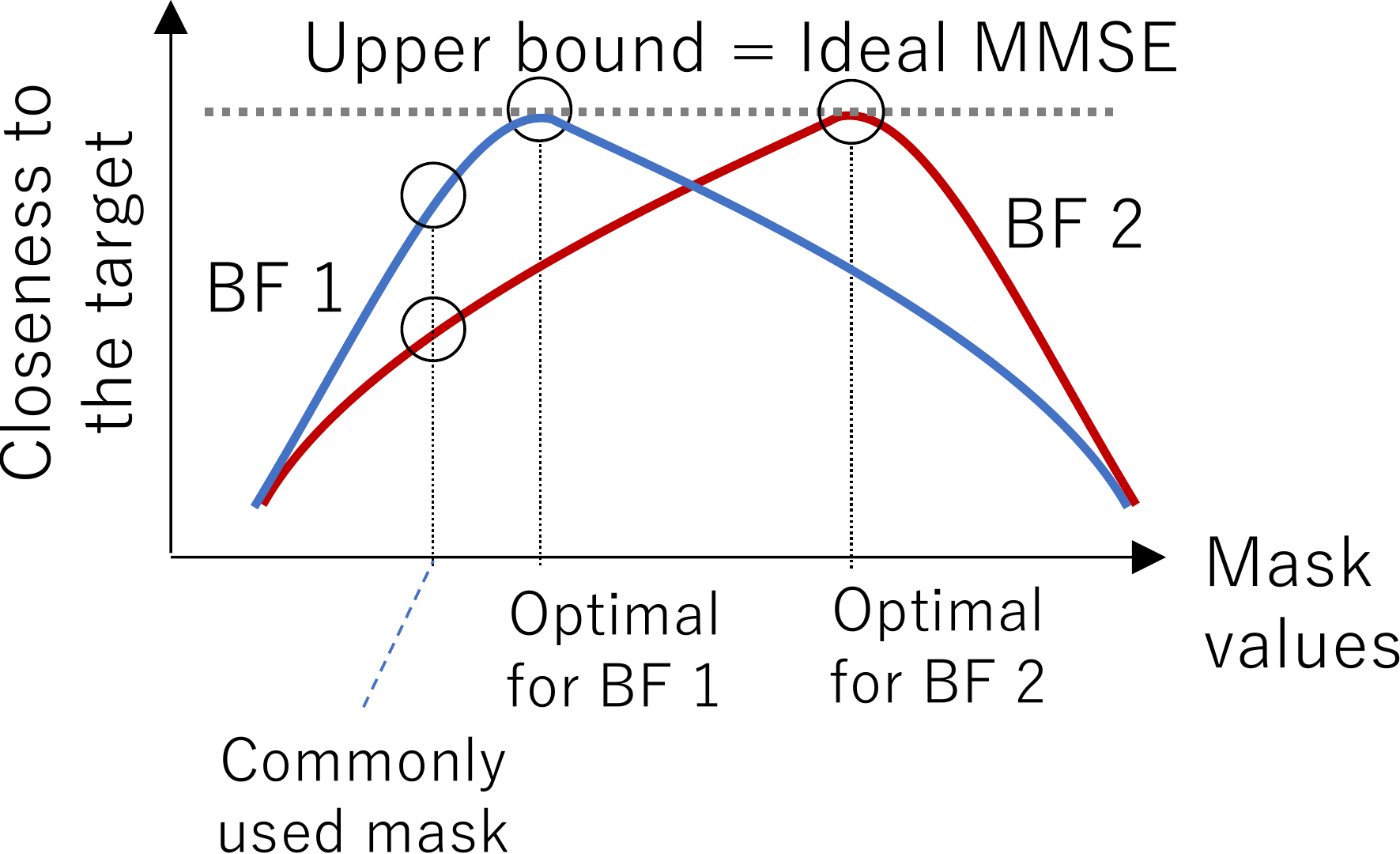}
  \caption{Schematic image of peak performance and optimal masks for two BFs; similar to \reffig{fig:mask_curve}, the vertical and horizontal axes indicate the closeness of the BF output to the target and mask values, respectively; a common mask may result in different extraction performances (closeness to the target) although the BFs have the same peak performance.}
  \label{fig:two-bf-case}
\end{figure}

\reffig{fig:two-bf-case} conceptually illustrates that multiple BFs contain the same peak extraction performance. The optimal mask differs for each BF, as illustrated in \reffig{fig:optimal-masks}. In the comparative studies~\cite{Heymann2016-sy,Boeddeker2018-ww,Wang2018-tl,Shimada2019-xj,Heymann2018-mf}, multiple BFs, such as BFs 1 and 2, used the same mask and demonstrated different performance scores. Although BF 1 appears to outperform BF 2 in \reffig{fig:two-bf-case}, this result does not contradict the fact that the peak performance is the same as BF 2. 

Another reason for the performance differences in these comparative studies is the inconsistent scaling methods employed, as mentioned in \refsubsec{subsec:scaling-methods}. For example, in~\cite{Heymann2016-sy}, the max-SNR BF was evaluated with BAN, while the MVDR BF was evaluated without scaling. Differences in scaling methods can significantly influence extraction performance even when the extraction filter is the same, as suggested in Experiment 3. Therefore, the scaling method needs to be unified for a fair comparison.

Furthermore, the differences in convergence speed verified in Experiment 1 can be a reason for the performance differences. \reffig{fig:rel-iteration} indicates that the INV-NS, corresponding to the Souden MVDR BF, converges the fastest, whereas ISEV-OS, corresponding to the MPDR BF, converges the slowest. These trends apply to the case that DNNs for mask estimation are jointly trained with the downstream tasks including the BF~\cite{Xu2019-um,Nguyen2022-hg,Heymann2017-vr}. Thus, different BFs may result in different performance scores if iterations are limited in the joint training.

\subsubsection{Concept of practical parameter count and performance saturation point}
The concept of the bias-variance tradeoff~\cite{Belkin2019-wv} can account for the results that all variations achieved the same peak performance as the theoretical upper-bound performance. The tradeoff implies that a model with many parameters can reduce an error (bias) between the model output and the supervisory data, but may increase the error (variance) between the output and unseen data, and vice versa for a model with fewer parameters. Viewing \refeqn{eqn:optimal-mask-def} and \refeqn{eqn:joint-optimization}, we can interpret that the mask-based BFs represent the problem of approximating the target (supervisory data) by employing one or two masks as a model parameter set. Given that the masks are optimized for each target in this study, we do not need to consider unseen data or increasing variance. 

The mask type categorization illustrated in \reffig{fig:mask-categorization} can be represented as differences in practical parameter count. A ratio mask contains more parameters than a binary mask but fewer parameters than a non-negative mask, considering that any non-negative mask can be decomposed into the maximum value and a ratio mask. Moreover, a complex-valued mask contains more parameters than two non-negative masks because it can be represented as two real-valued masks corresponding to the real and imaginary parts, and a real value can be decomposed into a sign and a non-negative value.

Significant assumptions include the bias represented as the approximation error in \refeqn{eqn:optimal-mask-def} and \refeqn{eqn:joint-optimization}, is determined solely by the practical parameter count, implying that variation types (e.g., MinGEV, INV, and ISEV) do not affect the bias; and that the BF output $z(t)$ minimizing the bias is uniquely determined independent of the BF variation used. The bias does not decrease further if the parameter count exceeds a particular number called the {\it saturation point}.

\begin{figure}[t]
  \centering
  \includegraphics[width=0.4\linewidth]{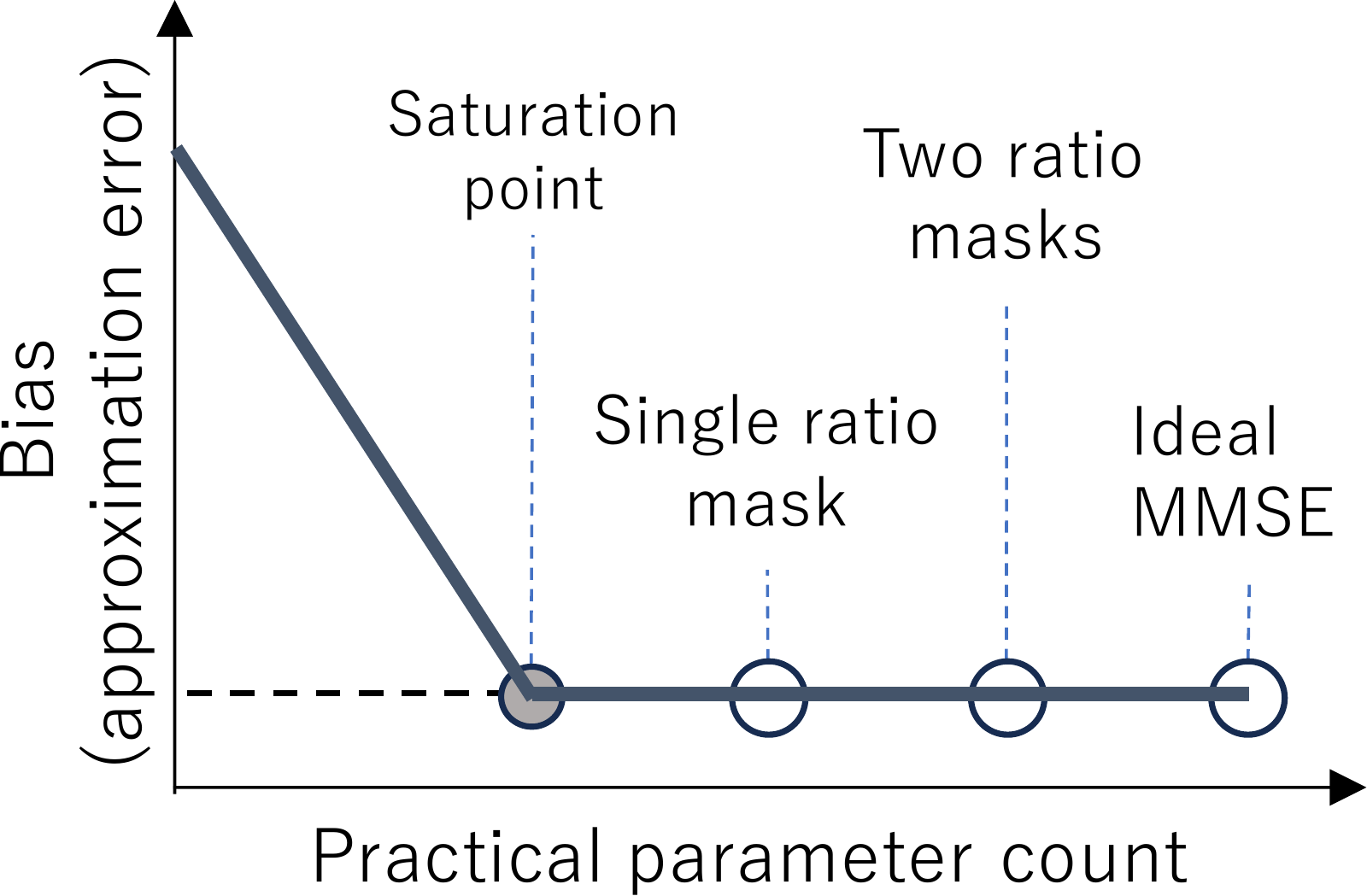}
  \caption{Schematic image of relationship between bias (approximation error) and practical parameter count for all BF variations and the ideal MMSE BF; a mask-based BF can be regarded as a model with parameters that approximate the target, and the mask count and type used affect the practical parameter count; we consider that all variations can achieve the same extraction performance as the ideal MMSE BF because these contain parameters more than the performance saturation point regardless of the mask count.}
  \label{fig:rel-bias-filter-estimation}
\end{figure}

We illustrate the relationship between bias and practical parameter count in \reffig{fig:rel-bias-filter-estimation}. The horizontal axis indicates relative count. Variations using two ratio masks, such as the MaxGEV-, MinGEV-, INV-, and ISEV-NS BFs in \reftable{tab:12variations}, are represented as a point labeled {\it two ratio masks}, whereas those using a single ratio mask are labeled {\it single ratio mask}. The former has twice as many parameters as the latter. The ideal MMSE BF includes the largest parameter count because it can be interpreted as a particular case using a complex-valued mask, as mentioned in \refsubsec{subsec:filter-estimation} and \refsubsec{subsec:mmse}. Given that all BF variations achieved the theoretical upper-bound performance obtained with the ideal MMSE BF, even variations using a single ratio mask exceed the saturation point.

An open question remains whether all variations exceed the saturation point for any dataset. Therefore, exploring peak extraction performance using various datasets is required.

\subsubsection{Uniqueness of the optimal mask}
\label{subsubsec:mask-unique}

\reffig{fig:optimal-masks} indicates that the optimal mask in filter estimation differs for each BF variation. We discuss the results in the following aspects:
\begin{enumerate}
\item Why do the optimal masks differ?
\item Is the optimal mask unique or multiple for each variation?
\end{enumerate}

For the first aspect, the reason is that the optimal mask is the solution to \refeqn{eqn:optimal-mask-def} or \refeqn{eqn:joint-optimization}, which represents a different minimization problem for each variation except for the equivalence between the MaxGEV and MinGEV types mentioned in \refsubsec{subsec:filter-estimation}. Therefore, each variation contains a different optimal mask except that the MaxGEV-NS, OS, and NO BFs contain the same optimal masks as the MinGEV-NS, OS, and NO BFs, respectively.

For the second aspect, the uniqueness of the optimal masks can be discussed as follows. As a common characteristic of all the variations, the optimal mask is scale-invariant because the mask scale (range of mask values) only affects those of $\vecw$ and $y(t)$ that can be adjusted in the scaling process. Therefore, the following masks are also optimal if $\ms(t)$ and $\mn(t)$ are optimal:
\begin{eqnarray}
\ms'(t) &=& a_{1}\ms(t), \label{eqn:scale-invariant-ms-prime}\\
\mn'(t) &=& a_{2}\mn(t), \label{eqn:scale-invariant-mn-prime}
\end{eqnarray}
where $a_{1}$ and $a_{2}$ denote arbitrary non-negative constants.

Additionally, the variations belonging to the MaxGEV and MinGEV types contain multiple optimal masks different from \refeqn{eqn:scale-invariant-ms-prime} and \refeqn{eqn:scale-invariant-mn-prime}, as proven in Appendix \ref{sec:generating-multiple-optimal-masks}. Therefore, an infinite number of optimal masks can be generated from one using the conversion rules shown in \reftable{tab:multiple-optimal-masks}, where the type prefixes such as MaxGEV- and MinGEV- are omitted considering the equivalence of the two types; for example, the variation named {\it NS} represents both the MaxGEV- and MinGEV-NS BFs. In the formulas of this table, $b_{1}$ and $b_{2}$ can be negative if $\ms'(t), \mn'(t)\geq 0$ for all $t$, whereas $a_{1}$ and $a_{2}$ must be non-negative. The bottom two rows in \reftable{tab:multiple-optimal-masks} indicate that the optimal mask for the MaxGEV- and MinGEV-OS BFs can be converted to the optimal masks for the other variations belonging to the MaxGEV and MinGEV types, and the same for the MaxGEV- and MinGEV-NO BFs.

However, the mask conversion rules shown in \reftable{tab:multiple-optimal-masks} do not apply to the variations belonging to the INV and ISEV types. Therefore, discussing whether these types contain multiple optimal masks is an open question except for the cases of \refeqn{eqn:scale-invariant-ms-prime} and \refeqn{eqn:scale-invariant-mn-prime}.

\begin{table}[b]
  \centering
\caption{Rules that generate multiple optimal masks from one in each variation belonging to the MaxGEV and MinGEV types ($\ms(t), \mn(t)$: optimal masks for each variation; $a_{1}, a_{2}$: arbitrary non-negative constants; $b_{1}, b_{2}$: arbitrary real-valued constants); Type prefixes, MaxGEV- and MinGEV-, are omitted in variation name; $\ms'(t)$ and $\mn'(t)$ are constrained such that $\ms'(t), \mn'(t)\geq 0$ for all $t$.}
\label{tab:multiple-optimal-masks}
\begin{tabular*}{0.7\textwidth}{@{\extracolsep\fill}ll}
\toprule
Variation name & Formula\\
\midrule\midrule
NS & $\ms'(t)=a_{1}\ms(t)+b_{1}\mn(t)$\\
   & $\mn'(t)=a_{2}\mn(t)+b_{2}\ms(t)$\\
\midrule
OS & $\ms'(t)=a_{1}\ms(t)+b_{1}$\\
\midrule
NO & $\mn'(t)=a_{2}\mn(t)+b_{2}$\\
\midrule\midrule
NO and NS from OS & $\mn'(t)=b_{2}-a_{2}\ms(t)$\\
\midrule
OS and NS from NO & $\ms'(t)=b_{1}-a_{1}\mn(t)$\\
\bottomrule
\end{tabular*}
\end{table}

\subsection{Why can scaling using L1-MN and L2-MN masks behave as the IS?}

\begin{figure}[t]
  \centering
  \includegraphics[width=0.4\linewidth]{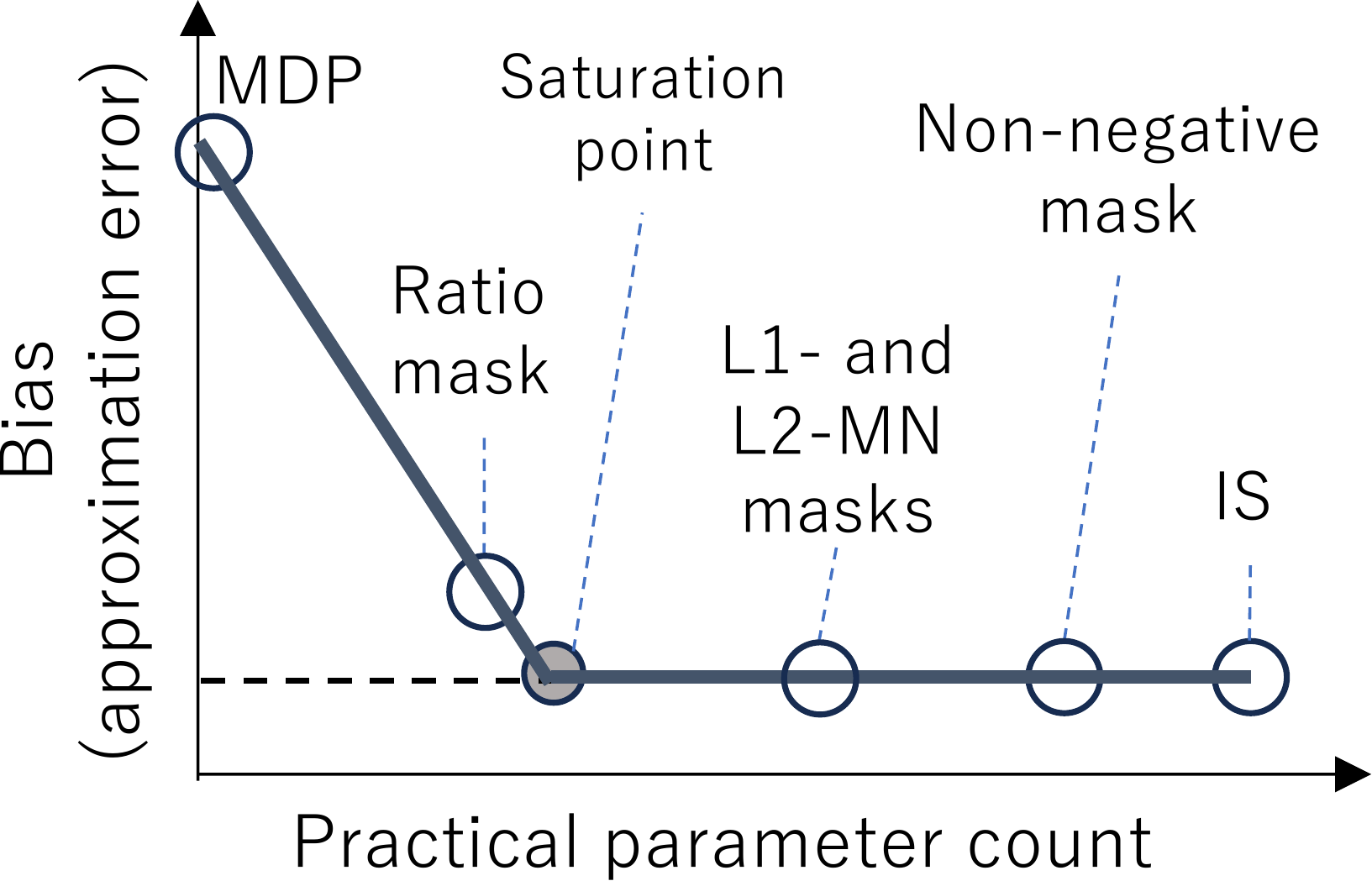}
  \caption{Schematic image of relationship between bias (approximation error) and practical parameter count for scaling methods; scaling can be regarded as approximating the target by using a model with parameters, and the mask type used affects the practical parameter count; we can consider that scaling methods using L1-MN, L2-MN, and non-negative masks contain more parameters, whereas that using a ratio mask contains slightly fewer parameters, compared with the saturation point.}
  \label{fig:rel-mask-scaling}
\end{figure}

Similar to the filter estimation, a scaling mask represented in \refeqn{eqn:optimal-scaling-mask-def} and \refeqn{eqn:joint-optimization} can also be interpreted as a model parameter set. We illustrate the relationship between bias and practical parameter count in \reffig{fig:rel-mask-scaling}. A non-negative mask contains more parameters than a ratio mask, as previously discussed. A family of MN masks such as L1-MN and L2-MN masks contain fewer parameters than a non-negative mask because these are constrained as \refeqn{eqn:l1-mean-normalized} and \refeqn{eqn:l2-mean-normalized}, although it is not evident whether the parameters are more than those of a ratio mask. The IS method contains the largest number of parameters because it corresponds to using a complex-valued mask, as mentioned in \refsubsec{subsec:scaling}. In contrast, the MDP includes no parameters because this corresponds to the case where all the mask values are fixed to 1.
 
The position of the saturation point is discussed. \reftable{tab:exp3} suggests that the practical parameter count of the L1-MN, L2-MN, and non-negative masks exceed the saturation point because these masks achieve the same performance as the IS, whereas that of a ratio mask is close to, but slightly lower than, the saturation point because results using the ratio mask appear to degrade slightly compared with the IS. Therefore, the practical parameter count for each scaling method can conceptually be plotted as \reffig{fig:rel-mask-scaling}.

Considering that a scaling mask will be estimated with a DNN in future work, a stronger mask constraint is more desirable for efficient DNN training, as mentioned in \refsubsec{subsec:constraints}. Therefore, a family of MN masks such as L1-MN and L2-MN masks are the most appropriate for scaling. In other words, the mask-based scaling is comparable with IS and applicable to realistic scenarios if the DNN can learn the optimal L1-MN or L2-MN mask.

\subsection{Significance of all variations achieving the upper-bound performance}
\label{subsec:comparison-using-chime}
Experiment 4 suggests that all the BF variations can achieve the same upper-bound extraction performance. Comparing this with the scores of existing TSE methods, such as mask-based BFs and ICA-based ones, we can contain the following insights for designing a TSE system:
\begin{enumerate}
\item The peak extraction performance of the mask-based BFs is not determined by the BF variation used; All BFs can achieve the upper-bound performance, the score with the ideal MMSE BF, if the corresponding ratio mask is optimally obtained.
\item Scaling is significant for the extraction performance; the mask-based scaling can behave as IS if the corresponding L1- or L2-MN mask is obtained.
\item A proper nonlinear post-process (NLPP) is required to outperform the upper-bound performance.
\end{enumerate}

We explain these aspects as follows. \reftable{tab:compare-tse} presents the SDR, PESQ, and STOI scores reported in the studies using the CHiME-4 (or CHiME-3) test sets. The eSTOI scores are omitted as they were only reported in~\cite{Chen2018-rl} and~\cite{Cho2021-ru}. For fair comparisons, the scores obtained with the batch (or offline) algorithms are chosen, although several studies also reported scores using online algorithms. Therefore, these scores may not be the best in each study. The column labeled NLPP indicates processes applied to the estimated target or BF output except for the scaling process represented as \refeqn{eqn:scaling-def}. Scores outperforming the ideal MMSE BF are underlined.

\begin{table}[b]
  \centering
\caption{Comparing TSE methods using CHiME-4 test set; scores outperforming the ideal MMSE BF are underlined. (NLPP: nonlinear post-process, FE: filter estimation)}
\label{tab:compare-tse}
\small
\begin{tabular*}{\textwidth}{@{\extracolsep\fill}llll|rrr}
\toprule
Method & FE & Scaling & NLPP & SDR & PESQ & STOI \\
       &    &         &      & [dB]&      & [\%] \\
\midrule\midrule
Ideal MMSE & - & - &-& 19.44 & 2.77 & 97.03 \\
\midrule
Heymann+16 \cite{Chen2018-rl,Heymann2016-sy} & MaxGEV-NS & - &-& 2.92 & 2.46 & 87\phantom{.00} \\
\multicolumn{1}{c}{(Max-SNR)}   &        &    &&&&\\
Erdogan+16 \cite{Erdogan2016-jc} & INV-NS    & - & - &  14.36 &\multicolumn{1}{c}{-}&\multicolumn{1}{c}{-}\\
\multicolumn{1}{c}{(Souden MVDR)} &&&&&&\\
Shimada+19 \cite{Shimada2019-xj} & INV-OS & - & - & 15.97 & 2.69 & 94\phantom{.00}\\
\multicolumn{1}{c}{(MMSE)} &&&&&&\\
\midrule
Cho+21 \cite{Cho2021-ru} (MLDR)& ISEV-NS & RTF &-&\multicolumn{1}{c}{-}& 2.70 &\multicolumn{1}{c}{-}\\
Hiroe 21 \cite{Hiroe2022-rd}  (SIBF)& MinGEV-NO & MDP &-& 17.29 & 2.72 & 96.18 \\
\midrule
Hiroe 21 \cite{Hiroe2022-rd} (SIBF) & MinGEV-NO & MDP & DNN & \underline{19.62} & \underline{3.00} & 96.36 \\
Wang+20 \cite{Wang2020-xk} (MVDR) & ISEV-NS & RTF & DNN & \underline{22.4}\phantom{0} & \underline{3.68} & \underline{98.6}\phantom{0} \\
\bottomrule
\end{tabular*}
\end{table}

The first row indicates the scores obtained with the ideal MMSE BF. As shown in \reftable{tab:exp4-test}, all BF variations achieve the same scores if the optimal masks are obtained. These scores only represent the upper-bound performance of linear TSE methods, leaving room for nonlinear methods to outperform.

The second to fourth rows indicate the scores of mask-based BFs, max-SNR~\cite{Chen2018-rl}, Souden MVDR~\cite{Erdogan2016-jc}, and MMSE BFs~\cite{Shimada2019-xj}, respectively. Each study discussed that the BF used caused the score differences. However, comparing the scores with \reftable{tab:exp4-test} suggests that the differences are due to the optimality of the mask and scale. Particularly, lower scores in~\cite{Chen2018-rl} were caused by omitting any scaling method although the max-SNR BF contains the scaling ambiguity issue as shown in \reftable{tab:formulas-mask-based-bfs}.

The fifth and sixth rows show the scores of the ICA-based TSE methods, MLDR BF~\cite{Cho2021-ru} and SIBF~\cite{Hiroe2022-rd}, respectively; the SIBF scores are the results of the sixth iterative casting mentioned in \refsubsec{subsec:sibf}. These methods compute $\mn(t)$ from the target source model based on different formulations than mask-based BFs, as mentioned in Appendices \ref{subsec:sibf} and \ref{subsec:mldr}. However, their upper-bound performance can be considered the same as the ideal MMSE BF, given that these correspond to the ISEV-NS and MinGEV-NO BFs shown in \reftable{tab:12variations}. Thus, we can discuss that the two methods obtain higher scores compared with the second to fourth rows because $\mn(t)$ associated with the target source model is closer to the optimal mask. Moreover, we can consider that proper target source models differ between MLDR and SIBF, given that the optimal masks differ between ISEV-NS and MinGEV-NO BFs as illustrated in \reffig{fig:optimal-masks}. Therefore, this study contributes to the ICA-based TSE methods, as well as the mask-based BFs.

Several studies reported scores outperforming the ideal MMSE BF despite using formulas included in the 12 variations~\cite{Hiroe2022-rd,Wang2020-xk}, as shown in the last two rows of \reftable{tab:compare-tse}. These results are attributed to the NLPP. In~\cite{Hiroe2022-rd}, scores were obtained by computing $r(t)z'(t)/|z'(t)|$, where $r(t)$ and $z'(t)$ denote the reference in the sixth iterative casting and SIBF output in the fifth casting, respectively. The NLPP of modifying the magnitude of the SIBF output with the reference-estimating DNN assisted in outperforming the upper-bound performance of the BFs. Meanwhile, in~\cite{Wang2020-xk}, a DNN for post-processing, different from the one for computing $\phis$ and $\phin$, was trained to estimate the target from both the MVDR output $z(t)$ and observation $x_{k}(t)$. We can estimate that the MVDR outputs underperformed the ideal MMSE BF, although the corresponding scores are not reported in~\cite{Wang2020-xk}; thus, the DNN-based post-process largely contributes to outperforming the ideal MMSE BF.

\subsection{Discussion on the INV-NO and ISEV-NO BFs}
\label{subsec:specific-discussion-on-several-variations}

The experimental results demonstrate that the INV-NO and ISEV-NO BFs achieved the theoretical upper-bound performance despite differing from conventional BFs. Here, we discuss the reason.

The formulas of INV-NO and ISEV-NO BFs contain $\phix\vece_{k}$ and $\sevmax{\phix}$, respectively, as shown in \reftable{tab:12variations}. These can be interpreted as inaccurate SVs compared with the MVDR (or ISEV-NS) BF formula represented in \refeqn{eqn:mvdr-filter}. However, this does not result in performance degradation unlike the MPDR BF mentioned in Appendix \ref{subsec:mvdr} because, 1)~using $\phin$ instead of $\phix$ prevents the target cancellation problem, and 2)~using mask-based scaling properly estimates the scale regardless of SV accuracy.

This discussion suggests that a novel variation involving both the INV-NO and ISEV-NO BFs can be proposed. However, it is estimated that this variation can also achieve the upper-bound performance if the practical parameter count exceeds the saturation point illustrated in \reffig{fig:rel-bias-filter-estimation}.

\section{Conclusions} \label{sec:conclusions}

This study explored the peak extraction performance of mask-based BFs. To compare multiple BFs under the same conditions, we proposed a unified framework for mask-based BFs consisting of two processes: filter extraction and scaling. To encompass all BF variations, we employed a classification rule based on the operators and covariance matrices within the formulas and identified 12 variations including two novel ones. These also covered ICA-based TSE methods like SIBF and MLDR BF. For the scaling process, we proposed a mask-based scaling method that can be combined with any BF variation and does not use the target. The optimal masks for both processes are obtained by minimizing the MSE between the target and BF output. We also examined the appropriate mask type for both processes, based on two perspectives: theoretical requirements in the formulation and constraints for efficient training of mask-estimating DNNs. Consequently, the framework allowed us to compare all possible BF variations under unified conditions. 

Through a series of experiments using the CHiME-4 dataset, where optimal masks were obtained utterance by utterance, we verified that; 1)~all 12 BF variations using ratio masks can commonly achieve theoretical upper-bound performance, 2)~mask-based scaling using an MN mask can act as the IS, and 3)~jointly optimizing both processes can also achieve the same performance.

In the discussion, we explained why the unified framework can achieve the upper-bound performance by considering the relationship between practical parameter count and saturation point, based on the bias-variance tradeoff concept. For filter extraction, all the variations are considered to surpass the saturation point in terms of parameter count, similar to the ideal MMSE BF. For scaling, the saturation point is considered to lie between the method using a ratio mask and that using an MN mask. This concept can account for the upper-bound performance of any novel variation proposed. We also indicated that the experimental results contribute to designing a TSE system by comparing the results with conventional studies using the same dataset. Finally, we discussed why several variations, such as the INV-NO and ISEV-NO BFs, can estimate the target despite being rarely employed as BFs. 

This study contributes to the following aspects:
\begin{enumerate}
    \item Designing a TSE system with higher extraction performance by indicating that extraction performance is determined not by the BF used, but by the mask estimation, scaling, and nonlinear post-processing. 
    \item Improving scaling accuracy by employing mask-based scaling. 
    \item Estimating the upper-bound performance of the BF used by employing the concept of the practical parameter count and saturation point. 
\end{enumerate}
These contributions also apply to ICA-based TSE methods because the unified framework includes formulas used in those methods.

Future work includes 1)~examining the extraction performance of the unified framework when masks are estimated with DNNs, and 2)~verifying that all BF variations can achieve the same peak performance by using other datasets.

The experimental system has been shared in \url{https://github.com/hreshare/unified_framework_for_mask-based_bf/}.

\bmhead{Abbreviations}
\begin{description}
\item[Abs:] absolute function
\item[AM:] acoustic model
\item[ASR:] automatic speech recognition
\item[BAN:] blind analytical normalization
\item[BF:] beamformer
\item[BN:] batch normalization
\item[dB:] decibel
\item[DNN:] deep neural network
\item[eSTOI:] extended short-time objective intelligibility measure
\item[GEV:] generalized eigenvalue decomposition
\item[ICA:] independent component analysis
\item[INV:] matrix inversion
\item[IS:] ideal scaling
\item[ISEV:] matrix inversion and standard eigenvalue decomposition
\item[JO:] joint optimization
\item[JT:] joint training
\item[L1-MN:] L1-mean-normalized
\item[L2-MN:] L2-mean-normalized
\item[MaxGEV:] maximum eigenvector in generalized eigenvalue decomposition
\item[max-ONR:] maximum observation-to-noise ratio
\item[max-SNR:] maximum signal-to-noise ratio
\item[max-SOR:] maximum signal-to-observation ratio
\item[MDP:] minimal distortion principle
\item[MinGEV:] minimum eigenvector in generalized eigenvalue decomposition
\item[min-NOR:] minimum noise-to-observation ratio
\item[min-NSR:] minimum noise-to-signal ratio
\item[min-OSR:] minimum observation-to-signal ratio
\item[MLDR:] maximum likelihood distortionless response
\item[MMSE:] minimum mean square error
\item[MN:] mean-normalized
\item[MPDR:] minimum power distortionless response
\item[MSE:] mean square error
\item[MVDR:] minimum variance distortionless response
\item[MWF:] multichannel Wiener filter
\item[NLPP:] nonlinear post-process
\item[NO:] $\phin$ and $\phix$
\item[NS:] $\phin$ and $\phis$
\item[OS:] $\phix$ and $\phis$
\item[PESQ:] perceptual evaluation of speech quality
\item[RTF:] relative transfer function
\item[SDR:] source-to-distortion ratio
\item[SEV:] standard eigenvector decomposition
\item[SIBF:] similarity-and-independence-aware BF
\item[STOI:] short-time objective intelligibility measure
\item[SV:] steering vector
\item[TF:] time-frequency
\item[TSE:] target sound extraction
\item[TV:] time-frequency-varying variance
\end{description}

\section*{Declarations}

\bmhead{Availability of data and materials}
This study used the CHiME-4 dataset. The availability of this dataset is described at \url{https://www.chimechallenge.org/challenges/chime4/data}.
\bmhead{Competing interests}
The authors declare that they have no competing interests.
\bmhead{Funding}
This work was supported by JST CREST
JPMJCR19K1.
\bmhead{Authors' contributions}
Conceptualization, A.H., K.I. and K.N.; methodology, A.H.; software, A.H.; validation, A.H.; formal analysis, A.H.; investigation, A.H.; writing-original draft preparation, A.H.; writing-review and editing, A.H. and K.N.; visualization, A.H.; supervision, K.I. and K.N.; project administration, A.H.; funding acquisition, K.I and K.N. All authors read and approved the published version of the manuscript.
\bmhead{Acknowledgements}
This work was supported by JST CREST JPMJCR19K1.
\bmhead{Code availability}
All source codes used in the experiments are available at 
\url{https://github.com/hreshare/unified_framework_for_mask-based_bf/}.

\begin{appendices}
\section{Formulation of existing linear TSE methods}
\label{sec:formulation-linear-tse}

This section describes the formulation of each linear TSE method including the mask-based BFs and ICA-based ones. The first three subsections explain the mask-based max-SNR, MMSE, and MVDR BFs. The remaining two subsections correspond to the ICA-based TSE such as the SIBF and MLDR BF.

\subsection{Max-SNR BF}
\label{subsec:max-snr}
The max-SNR BF group consists of six variations including the original one. The derivation of all variations is subsequently explained because this step is significant in enumerating all possible variations of the mask-based BFs and examining the uniqueness of the optimal mask.

The max-SNR (or MaxGEV-NS) BF is formulated as the following maximization problem~\cite{Heymann2016-eb,Heymann2016-sy}:
\begin{eqnarray}
\vecw
    &=& \argmax_{\vecw} \frac{\htp{\vecw}\phis\vecw}{\htp{\vecw}\phin\vecw}
            \label{eqn:max-snr-ratio} \\
    &=& \gevmax{\phis}{\phin}.
            \label{eqn:max-snr-gev}
\end{eqnarray}
Considering that both the numerator and denominator in \refeqn{eqn:max-snr-ratio} are nonnegative, \refeqn{eqn:max-snr-ratio} is equivalent to \refeqn{eqn:min-nsr-ratio}. Thus, we can obtain a variation called the minimum noise-to-signal ratio (min-NSR or MinGEV-NS) BF represented as \refeqn{eqn:min-nsr-gev}.
\begin{eqnarray}
\vecw
    &=& \argmin_{\vecw} \frac{\htp{\vecw}\phin\vecw}{\htp{\vecw}\phis\vecw}
            \label{eqn:min-nsr-ratio} \\
    &=& \gevmin{\phin}{\phis}.
            \label{eqn:min-nsr-gev}
\end{eqnarray}
Both the max-SNR and min-NSR BFs use two masks.
To derive the remaining variations that use a single mask, we assume the following relationship: 
\begin{equation}
     \phis + \phin = \phix .
        \label{eqn:cov_summation}
\end{equation}
This can eliminate $\phis$ in \refeqn{eqn:max-snr-ratio} to derive \refeqn{eqn:max-onr-ratio} and \refeqn{eqn:max-onr-gev}, referred to as the maximum observation-to-noise ratio (max-ONR or MaxGEV-NO) BF~\cite{Warsitz2007-fn}:
\begin{eqnarray}
\vecw
    &=& \argmax_{\vecw} \frac{\htp{\vecw}\phix\vecw}{\htp{\vecw}\phin\vecw}
            \label{eqn:max-onr-ratio} \\
    &=& \gevmax{\phix}{\phin}.
        \label{eqn:max-onr-gev}
\end{eqnarray}
Employing the equivalence between \refeqn{eqn:max-onr-ratio} and \refeqn{eqn:min-nor-ratio}, we can derive \refeqn{eqn:min-nor-gev} referred to as the minimum noise-to-observation ratio (min-NOR or MinGEV-NO) BF~\cite{Hiroe2022-rd,Hiroe2023-qh}:
\begin{eqnarray}
\vecw
    &=& \argmin_{\vecw} \frac{\htp{\vecw}\phin\vecw}{\htp{\vecw}\phix\vecw}
            \label{eqn:min-nor-ratio} \\
    &=& \gevmin{\phin}{\phix}.
        \label{eqn:min-nor-gev}
\end{eqnarray}
Similarly, eliminating $\phin$ in \refeqn{eqn:min-nsr-ratio} derives both \refeqn{eqn:min-osr-gev} and \refeqn{eqn:max-sor-gev}, referred to as the minimum observation-to-signal ratio (min-OSR or MinGEV-OS) and maximum signal-to-observation ratio (max-SOR or MaxGEV-OS)  BFs~\cite{Hiroe2023-qh}, respectively:
\begin{eqnarray}
\vecw
    &=& \argmin_{\vecw} \frac{\htp{\vecw}\phix\vecw}{\htp{\vecw}\phis\vecw}
            \label{eqn:min-osr-ratio} \\
    &=& \gevmin{\phix}{\phis},
            \label{eqn:min-osr-gev} \\
\vecw
    &=& \argmax_{\vecw} \frac{\htp{\vecw}\phis\vecw}{\htp{\vecw}\phix\vecw}
            \label{eqn:max-sor-ratio} \\
    &=& \gevmax{\phis}{\phix}.
            \label{eqn:max-sor-gev}
\end{eqnarray}

Constraints on the two masks included in $\phis$ and $\phin$ are considered. Given that both the numerator and denominator in \refeqn{eqn:max-snr-ratio} need to be nonnegative, both $\ms(t)$ and $\mn(t)$ also need to be nonnegative.

\subsection{MMSE BF}
\label{subsec:mmse}

The MMSE BF is formulated as a problem of minimizing the MSE between $y(t)$ and the given reference $q(t)$~\cite{Wang2018-tl,Malek2020-nw}:
\begin{eqnarray}
\vecw &=& \argmin_{\vecw} \average{\left|q(t) - y(t)\right|^{2}}_{t}
    \label{eqn:mmse-def} \\
        &=& \phix^{-1} \average{\vecx(t) \conj{q(t)}}_{t},
    \label{eqn:mmse-filter}
\end{eqnarray}
where $\conj{q(t)}$ denotes the conjugate of $q(t)$. In this study, we do not assume that $q(t)$ and $\vecn(t)$ are uncorrelated because $q(t)$ differs from $s_{k}(t)$. The mask-based MMSE BF employs the masked observation, which is $\ms(t) x_{k}(t)$, as the reference; thus, $\vecw$ can be obtained as
\begin{eqnarray}
\vecw &=& \argmin_{\vecw} \average{\left|\ms(t)x_{k}(t) - y(t)\right|^{2}}_{t}
    \label{eqn:mask-mmse-def} \\
                &=& \phix^{-1} \phis \vece_{k}.
    \label{eqn:mask-mmse}
\end{eqnarray}

Unlike the max-SNR BF, the formulation of the mask-based MMSE BF represented as \refeqn{eqn:mask-mmse-def} allows $\ms(t)$ to be any complex value.

\subsection{MVDR BF}
\label{subsec:mvdr}

The MVDR BF group consists of three variations. The minimum power distortionless response (MPDR) BF~\cite{Ehrenberg2010-tq} is included in this group.

The MPDR BF is formulated as the following minimization problem:
\begin{eqnarray}
\vecw &=& \argmin_{\vecw} \average{\left|y(t)\right|^{2}}_{t} \label{eqn:mpdr-formulation}\\
      && \mbox{s.t.}\ \htp{\vecw}\vech = 1 \label{eqn:distortionless-constraint} \\
      &=& \frac{\phix^{-1}\vech}{\htp{\vech}\phix^{-1}\vech}.
    \label{eqn:mpdr-filter}
\end{eqnarray}
This BF may suffer from the problem that the target is cancelled~\cite{Ehrenberg2010-tq} if the SV $\vech$ is inaccurately associated with the target direction.

In contrast, the MVDR can avoid the problem by employing $\phin$ instead of $\phix$ in \refeqn{eqn:mpdr-filter}~\cite{Ehrenberg2010-tq,Heymann2016-sy}:
\begin{eqnarray}
\vecw &=& \frac{\phin^{-1}\vech}{\htp{\vech}\phin^{-1}\vech}.
    \label{eqn:mvdr-filter}
\end{eqnarray}

A significant variation of the MVDR that does not employ an SV was proposed in~\cite{Souden2010-kp}, referred to as the {\it Souden MVDR}. This estimates projections of $y(t)$ to each microphone. The extraction filter for $k$th microphone can be obtained as
\begin{eqnarray}
\vecw &=& \frac{\phin^{-1}\phis\vece_{k}}{{\rm tr}\left(\phin^{-1}\phis\right)}. \label{eqn:souden-filter}
\end{eqnarray}
This BF can determine the scale of both $\vecw$ and $y(t)$ without any post-process.

We consider constraints on the two masks. Considering that $\vech$ is computed in \refeqn{eqn:steering-vector}, $\ms(t)$ can be any complex value. In contrast, $\mn(t)$ used in the MVDR must be non-negative, given that \refeqn{eqn:mvdr-filter} can be interpreted as \refeqn{eqn:mvdr-formulation} constrained with \refeqn{eqn:distortionless-constraint}, which is the problem of minimizing a weighted variance that needs to be non-negative.
\begin{eqnarray}
\vecw &=& \argmin_{\vecw} \average{\mn(t)\left|y(t)\right|^{2}}_{t}.
    \label{eqn:mvdr-formulation}
\end{eqnarray}
For the Sounden MVDR BF, however, we can consider that $\mn(t)$ becomes any complex value as mentioned in Appendix \ref{sec:trivial-optimal-masks-for-inv-ns}.

\subsection{SIBF}\label{subsec:sibf}

The SIBF is a method that extracts a source similar to a reference, which is an approximately estimated magnitude spectrogram of the target, leveraging not merely the mutual independence of the sources but also the dependence between the BF output and reference. This is formulated as the following minimization problem~\cite{Hiroe2022-rd}:
\begin{eqnarray}
\vecw &=& \argmin_{\vecw} \left\{-\log {\rm P}\bigl(y(t),r(t)\bigr)\right\} \label{eqn:sibf-formulation}\\
&& \mbox{s.t.}\quad \average{\left|y(t)\right|^{2}}_{t}=1, \label{eqn:sibf-constraint}
\end{eqnarray}
where a reference $r(t)$ denotes the magnitude spectrogram, while ${\rm P}\left(y(t),r(t)\right)$ represents a joint probability density function between the BF output and reference, referred to as a target source model. The reference can be generated with various TSE methods including DNN-based ones. An instance of a source model is the time-frequency-varying variance (TV) Gaussian model~\cite{Ramirez_Lopez2015-ib} written as
\begin{eqnarray}
{\rm P}\left(y(t),r(t)\right) &\propto& \exp\left(-\frac{\left|y(t)\right|^{2}}{\max\left(r(t)^{\beta},\varepsilon\right)}\right),
\label{eqn:tv-gauss}
\end{eqnarray}
where $\beta$ denotes a hyperparameter that controls the influence of the reference. The extraction filter for this model can be obtained as
\begin{equation}
\vecw = \gevmin{\average{\frac{\vecx(t)\htp{\vecx(t)}}{\max\left(r(t)^{\beta},\varepsilon\right)}}_{t}}{\phix}.
\label{eqn:sibf-tv-gauss}
\end{equation}

As discussed in~\cite{Hiroe2022-rd}, \refeqn{eqn:sibf-tv-gauss} corresponds to the min-NOR BF represented as \refeqn{eqn:min-nor-gev}, regarding $1/\max\left(r(t)^{\beta},\varepsilon\right)$ as $\mn(t)$ in \refeqn{eqn:phin}. Unlike the mask-based BFs, $\mn(t)$ is associated with not the interferences but the target.

Additionally, a technique of casting the SIBF output into the reference-estimating DNN was proposed in~\cite{Hiroe2022-rd}, referred to as {\it iterative casting}, to generate a more accurate reference and SIBF output. This technique also leads to the finding that the combination of the newer reference and the phase of the previous SIBF output tends to be more accurate than the newer SIBF output; in short, $r(t)z'(t)/|z'(t)|$ is better than both $z'(t)$ and $z(t)$, where $r(t)$ denotes the newer reference, while $z'(t)$ and $z(t)$ denotes the previous and newer SIBF outputs, respectively. 

\subsection{MLDR BF}\label{subsec:mldr}
The MLDR BF~\cite{Cho2019-pn,Cho2021-ru,Shin_undated-za} is formulated as a maximum likelihood estimation problem that estimates the extraction filter as follows:
\begin{eqnarray}
\phisigma &=& \average{\frac{\vecx(t)\htp{\vecx(t)}}{\sigma(t)^{2}}}_{t}, \label{eqn:phisigma}\\
\vecw &=& \frac{\phisigma^{-1}\vech}{\htp{\vech}\phisigma^{-1}\vech},
    \label{eqn:mldr-filter} \\
\sigma(t)^{2} &=& \left|\htp{\vecw}\vecx(t)\right|^2,
    \label{eqn:calc-tv-in-mldr}
\end{eqnarray}
where $\sigma(t)^{2}$ denotes a TV of the target based on the TV Gaussian model. Given that $\sigma(t)$ is also a parameter to be estimated, $\vecw$ and $\sigma(t)$ are alternatively computed by using \refeqn{eqn:phisigma} to \refeqn{eqn:calc-tv-in-mldr}. As a variation of the MLDR BF, $\sigma(t)$ is employed as the denominator of \refeqn{eqn:phisigma} in~\cite{Shin_undated-za}, based on the TV Laplacian model.

Comparing \refeqn{eqn:mldr-filter} and \refeqn{eqn:mvdr-filter}, the MLDR BF can correspond to the MVDR BF, regarding $1/\sigma(t)^{2}$ as $\mn(t)$. Similar to the SIBF, $\mn(t)$ is associated with the target.

\section{Trivial optimal masks for INV type}
\label{sec:trivial-optimal-masks-for-inv-ns}

The variations belonging to the INV type shown in \reftable{tab:12variations} contain the trivial optimal masks if mask values can be any complex numbers. The derivation is explained.

We assume that $\phin$ contains the inverse matrix and consider the following equations to find the trivial optimal masks for the INV-NS BF:
\begin{eqnarray}
\vecw_{\rm ideal} &=& \phin^{-1} \phis \vece_{k}\\
\Leftrightarrow
\phin \vecw_{\rm ideal} &=& \phis \vece_{k}, \label{eqn:inv-ns-ideal}
\end{eqnarray}
where $\vecw_{\rm ideal}$ denotes the extraction filter obtained with \refeqn{eqn:ideal-mmse}. A sufficient condition of \refeqn{eqn:inv-ns-ideal} is that \refeqn{eqn:inv-ns-suff-cond1} is satisfied for all $t$, and a sufficient condition of this is \refeqn{eqn:inv-ns-suff-cond2}.
\begin{eqnarray}
\mn(t) \vecx(t)\htp{\vecx(t)} \vecw_{\rm ideal} &=& \ms(t) \vecx(t)\htp{\vecx(t)} \vece_{k}, \label{eqn:inv-ns-suff-cond1}\\
\mn(t) \htp{\vecx(t)}\vecw_{\rm ideal} &=& \ms(t) \htp{\vecx(t)}\vece_{k}
\quad\left(= \ms(t)\conj{x_{k}(t)}\right). \label{eqn:inv-ns-suff-cond2}
\end{eqnarray}
The inverse matrix of $\phin$ does not exist if $\mn(t)=0$ for all $t$. Therefore, the trivial optimal masks, $\ms(t)$ and $\mn(t)$, can be represented as the following ratio:
\begin{eqnarray}
\frac{\ms(t)}{\mn(t)} &=& \frac{\htp{\vecx(t)}\vecw_{\rm ideal}}{\conj{x_{k}(t)}}.
\label{eqn:inv-ns-trivial-mask}
\end{eqnarray}

Similarly, the trivial optimal masks for the INV-OS and INV-NO BFs can be obtained as \refeqn{eqn:inv-os-trivial-mask} and \refeqn{eqn:inv-no-trivial-mask}, respectively.
\begin{eqnarray}
\ms(t) &=& \frac{\htp{\vecx(t)}\vecw_{\rm ideal}}{\conj{x_{k}(t)}},
\label{eqn:inv-os-trivial-mask}\\
\mn(t) &=& \frac{\conj{x_{k}(t)}}{\htp{\vecx(t)}\vecw_{\rm ideal}}.
\label{eqn:inv-no-trivial-mask}
\end{eqnarray}

The INV-OS BF contains another optimal mask represented as \refeqn{eqn:inv-os-trivial-mask2}, considering that this makes \refeqn{eqn:mask-mmse-def} identical to \refeqn{eqn:ideal-mmse}.
\begin{eqnarray}
\ms(t) &=& \frac{\;\conj{s_{k}(t)}\;}{\;\conj{x_{k}(t)}\;}.
\label{eqn:inv-os-trivial-mask2}
\end{eqnarray}

Note that the right-hand sides of \refeqn{eqn:inv-ns-trivial-mask}--\refeqn{eqn:inv-os-trivial-mask2} are complex-valued. Therefore, the non-negative and more constrained masks cannot satisfy these equations.

\section{Generating multiple optimal masks from one in the MaxGEV and MinGEV types}
\label{sec:generating-multiple-optimal-masks}

Each BF variation belonging to the MaxGEV and MinGEV types contains multiple optimal masks. We derive rules that generate different optimal masks from one. Hereafter, $\ms(t)$ and $\mn(t)$ denote the optimal masks for each variation.

For the MaxGEV-NS BF, the following masks are also optimal if both $\ms(t)$ and $\mn(t)$ are optimal:
\begin{eqnarray}
\ms'(t) &=& a_{1}\ms(t) + b_{1}\mn(t)\mbox{\quad s.t.\quad$\ms'(t)\geq 0$ for all $t$}, \label{eqn:maxgev-ns-ms-prime}\\
\mn'(t) &=& a_{2}\mn(t) + b_{2}\ms(t)\mbox{\quad s.t.\quad$\mn'(t)\geq 0$ for all $t$}, \label{eqn:maxgev-ns-mn-prime}
\end{eqnarray}
where $a_{1}$ and $a_{2}$ denote arbitrary non-negative constants, whereas $b_{1}$ and $b_{2}$ are arbitrary real-valued constants. Both $b_{1}$ and $b_{2}$ can be negative if both $\ms'(t)$ and $\mn'(t)$ are non-negative. The mask optimality can be proven by assigning $\ms(t)=(\ms'(t)-b_{1}\mn(t))/a_{1}$ and $\mn(t)=(\mn'(t)-b_{2}\ms(t))/a_{2}$ to \refeqn{eqn:max-snr-ratio} and \refeqn{eqn:min-nsr-ratio}, respectively, considering that this variation is based on the max-SNR BF formulation. These masks are also optimal for the MinGEV-NS BF because of the equivalence between both BFs.

For the MaxGEV-OS BF, $\ms'(t)$ calculated in \refeqn{eqn:maxgev-os-ms-prime} is also optimal if $\ms(t)$ is optimal.
\begin{eqnarray}
\ms'(t) &=& a_{1}\ms(t) + b_{1}\mbox{\quad s.t.\quad$\ms'(t)\geq 0$ for all $t$}. \label{eqn:maxgev-os-ms-prime}
\end{eqnarray}
This can be proven by assigning $\ms(t)=(\ms'(t)-b_{1})/a_{1}$ to \refeqn{eqn:max-sor-ratio}. This mask is also optimal for the MinGEV-OS BF because of the equivalence of both BFs. We can convert $\ms(t)$ to the optimal mask for the MaxGEV- and MinGEV-NO BFs as follows:
\begin{eqnarray}
\mn'(t) &=& b_{2} - a_{2}\ms(t) \mbox{\quad s.t.\quad$\mn'(t)\geq 0$ for all $t$}. \label{eqn:mingev-no-mn-prime}
\end{eqnarray}
This can be proven by the fact that assigning $\ms(t)=(b_{2}-\mn'(t))/a_{2}$ to \refeqn{eqn:max-sor-ratio} results in the same formula as \refeqn{eqn:min-nor-ratio}. Additionally, both $\mn'(t)$ calculated in \refeqn{eqn:mingev-no-mn-prime} and $\ms(t)$ can be employed as the optimal masks for the MaxGEV- and MinGEV-NS. This can be proven by the fact that replacing $\mn(t)$ with $\mn'(t)$ in \refeqn{eqn:min-nsr-ratio} results in the same formula as \refeqn{eqn:min-osr-ratio}.

Similarly, for the MaxGEV-NO and MinGEV-NO BFs, $\mn'(t)$ calculated in \refeqn{eqn:maxgev-no-mn-prime} is also optimal if $\mn(t)$ is optimal, and $\ms'(t)$ calculated in \refeqn{eqn:mingev-os-ms-prime} can be employed as the optimal mask for the MaxGEV- and MinGEV-OS BFs in contrast to \refeqn{eqn:mingev-no-mn-prime}.
\begin{eqnarray}
\mn'(t) &=& a_{2}\mn(t) + b_{2}\mbox{\quad s.t.\quad$\mn'(t)\geq 0$ for all $t$}, \label{eqn:maxgev-no-mn-prime}\\
\ms'(t) &=& b_{1} - a_{1}\mn(t) \mbox{\quad s.t.\quad$\ms'(t)\geq 0$ for all $t$}. \label{eqn:mingev-os-ms-prime}
\end{eqnarray}
Additionally, both $\mn(t)$ and $\ms'(t)$ calculated in \refeqn{eqn:mingev-os-ms-prime} can be employed as the optimal masks for the MaxGEV- and MinGEV-NS.

\end{appendices}


\bibliography{paperpile}

\end{document}